\documentclass[acmsmall]{acmart}
\usepackage{type1cm} 
\usepackage{graphicx} 
\usepackage{xspace} 
\usepackage{balance} 
\usepackage{booktabs} 
\usepackage{multirow} 
\usepackage[font={bf}, tableposition=top]{caption} 
\usepackage{bold-extra} 
\usepackage{siunitx} 
\usepackage[vlined,linesnumbered,ruled,noend]{algorithm2e} 
\usepackage{microtype} 
\usepackage{xfrac} 
\usepackage{mathtools} 
\PassOptionsToPackage{hyphens}{url} 
\PassOptionsToPackage{bookmarks, pdftex, colorlinks=true, pagebackref=true, backref=page}{hyperref} 
\usepackage{cleveref} 
\PassOptionsToPackage{square,numbers}{natbib} 
\usepackage[hyperpageref]{backref} 
\usepackage{hyphenat} 
\usepackage{subcaption} 
\usepackage[export]{adjustbox} 

\renewcommand*\backref[1]{\ifx#1\relax \else (Cited on #1) \fi}

\setcopyright{acmlicensed}
\copyrightyear{2025}
\acmYear{2025}
\acmDOI{XXXXXXX.XXXXXXX}

\usepackage{multirow}
\usepackage{graphicx}
\usepackage{enumitem}
\usepackage[all]{nowidow}
\usepackage{longtable}

\graphicspath{{img/}}

\title{Language-Agnostic Modeling of Source Reliability on Wikipedia}

\author{Jacopo D'Ignazi}
\affiliation{%
  \institution{Universitat Pompeu Fabra}
  \city{Barcelona}
  \country{Spain}
}
\affiliation{%
  \institution{ISI Foundation}
  \city{Turin}
  \country{Italy}
}
\author{Andreas Kaltenbrunner}
\affiliation{%
  \institution{Universitat Oberta de Catalunya}
  \city{Barcelona}
  \country{Spain}
}
\affiliation{%
  \institution{ISI Foundation}
  \city{Turin}
  \country{Italy}
}
\affiliation{%
  \institution{Universitat Pompeu Fabra}
  \city{Barcelona}
  \country{Spain}
}

\author{Yelena Mejova}

\author{Kyriaki Kalimeri}
\affiliation{%
  \institution{ISI Foundation}
  \city{Turin}
  \country{Italy}
}
\author{Michele Tizzani}
\affiliation{%
  \institution{ISI Foundation}
  \city{Turin}
  \country{Italy}
}
\affiliation{%
  \institution{Technical University of Denmark}
  \city{Copenhagen}
  \country{Denmark}
}
\author{Mariano G. Beiró}
\affiliation{%
  \institution{Universidad de San Andrés}
  \city{Buenos Aires}
  \country{Argentina}
}
\affiliation{%
  \institution{CONICET}
  \city{Buenos Aires}
  \country{Argentina}
}
\author{Pablo Aragón}
\affiliation{%
  \institution{Universitat Pompeu Fabra}
  \city{Barcelona}
  \country{Spain}
}

\renewcommand{\shortauthors}{D'Ignazi et al.}

\newcommand{\MBremoved}[1]{}

\begin{abstract}
Over the last few years, verifying the credibility of information sources has become a fundamental need to combat disinformation.
Here, we present a language-agnostic model designed to assess the reliability of web domains as sources in references across multiple language editions of Wikipedia.
Utilizing editing activity data, the model evaluates domain reliability within different articles of varying controversiality, such as Climate Change, COVID-19, History, Media, and Biology topics.
Crafting features that express domain usage across articles, the model effectively predicts domain reliability, achieving an F1 Macro score of approximately 0.80 for English and other high-resource languages.
For mid-resource languages, we achieve 0.65, while the performance of low-resource languages varies. In all cases, the time the domain remains present in the articles (which we dub as \emph{permanence}) is one of the most predictive features.
We highlight the challenge of maintaining consistent model performance across languages of varying resource levels and demonstrate that adapting models from higher-resource languages can improve performance.
We believe these findings can assist Wikipedia editors in their ongoing efforts to verify citations and may offer useful insights for other user-generated content communities.
\end{abstract}


\begin{document}


\maketitle

\section{Introduction}
\label{sec:intro}
The proliferation of fake news in recent years is deteriorating the reliability and trustworthiness of the online information ecosystem~\cite{del2016spreading,vosoughi2018spread}. To mitigate their spread and impact in society, researchers in web and data mining are devoting a great deal of effort to generating resources to detect untrue content on social media platforms~\cite{su2023mining,xu2022identifying,ashkinaze2024dynamics,xiao2024msynfd,guo2023interpretable,xiao2023hipo,wu2023prompt}.

In popular web platforms like Wikipedia\footnote{Wikipedia is among the top visited websites worldwide:~\url{https://www.semrush.com/trending-websites/global/all}.}, disinformation is also a main threat to knowledge integrity~\cite{aragon2021preliminary}.  Wikipedia's conception of knowledge as a service requires content to be verifiable to preserve its reliability and integrity\footnote{\url{https://research.wikimedia.org/knowledge-integrity.html}}. Thus, the platform's policy on verifiability implies that readers should be able to check that all the information in Wikipedia articles comes from reliable sources\footnote{\url{https://en.wikipedia.org/wiki/Wikipedia:Verifiability.}}. While a classical patrolling technique in this platform is to detect and remove statements that violate basic core content policies~\cite{kumar2016disinformation}, some Wikipedia editors have proven more effective in tackling disinformation by first identifying unreliable sources~\cite{wired2021one}. Recent research revealed the positive impact of the community-curated list of perennial sources in English Wikipedia~\cite{Baigutanova_2023}. However, that list is limited or even non-existent in other language editions~\cite{baigutanova2023comparative}. A great deal of manual effort would be necessary to create such community-curated resources in all of the languages on Wikipedia.

To address this challenge, we propose a language-agnostic approach to reveal source reliability that leverages the implicit signals in the edit activity data from multiple Wikipedia language editions.
We aim to support Wikipedia editors, including those in
medium and small projects that often miss advanced tools, in identifying sources with patterns associated with low reliability, and in monitoring their prevalence across language editions. To meet this goal, we address the following research questions:

\begin{itemize}

\item \textbf{RQ1}. What are the most predictive language-agnostic features when modeling source reliability in Wikipedia, considering the Climate Change topic in English as a case study?

\item \textbf{RQ2}. 
 To what extent can model performance be related to the topic and size of a language edition of Wikipedia?

\item \textbf{RQ3}. 
Is it possible to adapt these models across topics and/or languages?

\end{itemize}

Our methods to answer these questions are inspired by
existing 
language-agnostic approaches to measure the controversiality of linked elements in a given Wikipedia article~\cite{borra2015contropedia}. Results with datasets from different topics and language editions suggest a promising scenario.
Given the several limitations 
outlined in the following sections, we pay attention to model performance not only for the largest editions but also in mid- and low-resource settings, where adaptation of the models may provide the most benefit.
To our knowledge, this is the first work 
modeling 
source reliability on Wikipedia using only language-agnostic features.

\section{Related Work}
\label{sec:relwork}

This work approaches the challenge of measuring source reliability on Wikipedia through the lens of controversiality metrics. This approach builds on the methodology of the Contropedia project~\cite{borra2015contropedia}, aimed at identifying and visualizing controversial links between Wikipedia articles, i.e., links that point to other articles in the platform and whose collaborative edit history shows signs of dispute between editors before reaching consensus.  This is not the only effort in this space. Systems like Edit-History Vis~\cite{guo2023edit} have also included interactive visualizations to illustrate the path to consensus for a Wikipedia article. Cultural differences in the controversiality of specific topics have been shown to exist (e.g.,~\cite{rogers2012neutral,fu2023introducing}). Unlike these previous approaches primarily focusing on wikilinks within article text, our work shifts attention to citations, aiming to explore the reliability of sources in Wikipedia.

Previous research has examined assessed the verifiability of Wikipedia content~\cite{chen2012citation,benjakob2022citation,halitaj2024providing,halitaj2025alpet} and the reliability of the sources~\cite{lewoniewski2020modeling, yang2024polarization}. ~\citet{Baigutanova_2023} addressed both goals by analyzing reference quality on English Wikipedia over time using two key metrics: the proportion of sentences lacking necessary citations, as identified by machine learning models~\cite{redi2019citation,chou2020citation}, and the proportion of citations linking to non-authoritative sources. For the latter, non-authoritative sources were determined using the \textit{perennial source list}~\cite{perennial_sources}, a curated list of sources whose reliability and use on Wikipedia are frequently discussed. The study found the share of references to non-authoritative sources has remained below 1\%, with a notable decline following the introduction of the list in 2018. 

Although the \textit{perennial source list} proved effective in decreasing the presence of non-authoritative sources on English Wikipedia, later research revealed significant limitations when applying this concept to other language editions~\cite{baigutanova2023comparative}. A few other language communities have developed their own versions of the \textit{perennial source list}, but such cases are rare among the hundreds of Wikipedia editions, and these lists typically contain only limited data. Furthermore, sources flagged as unreliable in one language edition are often still used in others, highlighting the variability in how source reliability is perceived across cultural and linguistic contexts. These discrepancies pose a major challenge to the development of universally applicable models for assessing source reliability across Wikipedia language editions.


The multilingual nature of Wikipedia calls for specific modeling techniques~\cite{johnson2022considerations}. Models such as multilingual BERT have been used to detect vandalism~\cite{trokhymovych2021wikicheck} or political bias~\cite{swati2023commonsense}, while other works have built rankings of relative quality in multiple language versions of the platform~\cite{lewoniewski2020modeling,lewoniewski2019multilingual}.
To build a multilingual model, such as one recently proposed for entity insertion \cite{feith2024entity}, one needs a large dataset of training data in all the languages of interest.
For some other tasks, and due to the limitations found in languages with smaller Wikipedia datasets, \emph{language-agnostic} approaches. These approaches have proven effective across a wide range of tasks, e.g., the assessment of article content quality~\cite{das2024language}, reliability~\cite{wong2021wiki}, automatic entity-linking~\cite{gerlach2021multilingual}, orphan article detection~\cite{arora2024orphan}, and topic modeling~\cite{piccardi2021crosslingual} and classification~\cite{johnson2021language}.

Language-agnostic models using features extracted from the network structure or user behavior have also been used in the more general context of detecting misinformation or low-quality information in social media. In this line,~\citet{shu2019studying} constructed embeddings of users and news and used diffusion models to trace the paths through which fake news propagates, while~\citet{zhao2021detecting} extracted features from the user interaction network to help identify misinformation in online health communities. As shown later, user-behavior related language-agnostic signals are useful for modeling source reliability, and can be adapted across languages to improve performance in low-resource settings.

\section{Data and Methods}
\label{sec:data}


\subsection{Perennial Sources}

We use the English Wikipedia perennial source list~\cite{perennial_sources} as ground truth for modeling source reliability.
As only a few other languages have perennial source lists, in this work we focus on the list from English Wikipedia, as it is the most complete one, and concerns the largest language Wikipedia. We discuss this lack of ground truth data in the Limitations section below.
It was collected on March 30, 2023. 
Although the list exists for 12 other languages, we only use the English version as it is by far the most extensive (see~\cite{baigutanova2023comparative} for details on coverage).
In English, it consists of a collection of web domains in five categories: blacklisted (569), deprecated (125), generally unreliable (178), no consensus (137), and generally reliable (184). 
To build a binary classifier, we use the generally reliable domains as positive labels, merge the domains blacklisted, deprecated, and generally unreliable into a single unreliable category (i.e., the negative label), and ignore the no consensus domains.

\begin{table*}[!t]
\caption{Statistics of the topic-language datasets used in the study grouped by high, mid, and low resource languages. All numbers are averages (apart from the number of languages). Perennial Domains (P. Domains) refers to the number of domains for the perennial source list in English found in the different datasets. See Table~\ref{tab:dataset_stats_median} in Appendix~\ref{app:medians} for the corresponding median values.}
\label{tab:dataset_stats}
\centering
\footnotesize
\begin{tabular}{@{}clrrrrrrrrr@{}}
\toprule
& \textbf{Topic} & \textbf{Langs} & \textbf{Articles} & \textbf{Revs} & \textbf{Revs/Ar} & \textbf{URLs} & \textbf{URLs/Ar} & \textbf{Domains} & \textbf{P. Domains} \\
\midrule
\multirow{5}{*}{\rotatebox[origin=c]{90}{High}}
&Climate Change & 7 & $\num{1378}$ & $\num{372133}$ & $\num{269}$ & $\num{58671}$ & $\num{49}$ & $\num{15363}$ & $\num{203}$ \\
&COVID-19 & 7 & $\num{898}$ & $\num{233905}$ & $\num{260}$ & $\num{60168}$ & $\num{89}$ & $\num{8289}$ & $\num{206}$ \\
&Biology Sample & 7 & $\num{13607}$ & $\num{586171}$ & $\num{43}$ & $\num{59763}$ & $\num{6}$ & $\num{11713}$ & $\num{148}$ \\
&History Sample & 7 & $\num{5710}$ & $\num{624428}$ & $\num{109}$ & $\num{52702}$ & $\num{12}$ & $\num{12121}$ & $\num{166}$ \\
&Media Sample & 7 & $\num{5577}$ & $\num{819541}$ & $\num{146}$ & $\num{108583}$ & $\num{22}$ & $\num{21290}$ & $\num{259}$ \\
\midrule
\multirow{5}{*}{\rotatebox[origin=c]{90}{Mid}}
&Climate Change & 36 & $\num{395}$ & $\num{28364}$ & $\num{71}$ & $\num{6224}$ & $\num{19}$ & $\num{2599}$ & $\num{94}$ \\
&COVID-19       & 37 & $\num{272}$ & $\num{18523}$ & $\num{68}$ & $\num{8984}$ & $\num{39}$ & $\num{1788}$ & $\num{102}$ \\
&Biology & 37 & $\num{4153}$ & $\num{83599}$ & $\num{20}$ & $\num{8691}$ & $\num{3}$ & $\num{2022}$ & $\num{61}$ \\
&History & 37 & $\num{1680}$ & $\num{83258}$ & $\num{49}$ & $\num{6274}$ & $\num{7}$ & $\num{2054}$ & $\num{67}$ \\
&Media   & 37 & $\num{1018}$ & $\num{46957}$ & $\num{46}$ & $\num{7932}$ & $\num{10}$ & $\num{2416}$ & $\num{123}$ \\
\midrule
\multirow{5}{*}{\rotatebox[origin=c]{90}{Low}}
&Climate Change & 56 & $\num{115}$ & $\num{3430}$ & $\num{29}$ & $\num{1079}$ & $\num{12}$ & $\num{570}$ & $\num{35}$ \\
&COVID-19       & 75 & $\num{32}$ & $\num{916}$ & $\num{28}$ & $\num{568}$ & $\num{22}$ & $\num{242}$ & $\num{32}$ \\
&Biology & 41 & $\num{593}$ & $\num{21096}$ & $\num{35}$ & $\num{1159}$ & $\num{4}$ & $\num{460}$ & $\num{16}$ \\
&History & 55 & $\num{511}$ & $\num{13501}$ & $\num{26}$ & $\num{952}$ & $\num{6}$ & $\num{406}$ & $\num{22}$ \\
&Media   & 70 & $\num{159}$ & $\num{2262}$ & $\num{14}$ & $\num{528}$ & $\num{7}$ & $\num{264}$ & $\num{30}$ \\
\bottomrule
\end{tabular}
\end{table*}

\subsection{Language Groups}

As we aim to test the proposed model on a variety of languages, we collect data on all $326$ available language editions of Wikipedia.
We then discard all languages that have fewer than two reliable and unreliable sources according to the English perennial source list.
Among the remaining ones, we distinguish between low, middle, and high-resource languages by examining the number of ``active users'', defined as ``registered users who have made at least one edit in the last $30$ days.''\footnote{\url{https://meta.wikimedia.org/wiki/List_of_Wikipedias} (accessed on 01-03-2024)}
We consider the top 5\% of the languages having the most active users as high-resource, the subsequent 25\% as mid-resource, and the remaining 70\% as low-resource languages.

\subsection{Article Collection}

\subsubsection{Topic Selection}

In this study, we have focused on articles from the following five topic datasets: Climate Change, COVID-19, Biology, History, and Media.
Climate Change and COVID-19 were selected as they are both controversial, but for different time durations. COVID-19 appeared at the end of 2019, while Climate Change has been a controversial topic for decades~\cite{korte2023causes}.
Both have a Wikiproject community of editors~\footnote{\url{https://en.wikipedia.org/wiki/Wikipedia:WikiProject}} who label articles as relevant to the topic with special guidelines.
For comparison, we also selected three topics that are not as controversial, but which varied in terms of reference quality
based on the ``reference risk'' metric. This metric was introduced by \citet{Baigutanova_2023} and estimates the likelihood of articles citing unreliable sources. Specifically, reference risk is calculated as the proportion of citations in an article that point to domains categorized as generally unreliable, deprecated, or blacklisted in the English Wikipedia perennial source list. Higher reference risk indicates a greater potential threat to verifiability and information quality.

This stratification allowed us to evaluate our model’s performance across topics with varying levels of citation quality and editorial scrutiny.

\subsubsection{Data Download}

For the first two topics, we use the articles listed by the corresponding topic-based WikiProjects.
For the latter three, we use the topics identified using the Wikipedia Objective Revision Evaluation Service (ORES) machine learning system~\cite{halfaker2020ores}. We use this method rather than later approaches~\cite{johnson2021language}, since it is the one employed in \citet{Baigutanova_2023}, which inspires our comparison among topics.
To keep the sizes of datasets comparable and manageable, we sample the articles identified by ORES to be in Biology and History topics by 10\% (each), and in the Media topic by 2\%.
These two approaches allow us to find articles in the English Wikipedia for each of the five topics, resulting in a total of \num{84027} articles.
To find the versions of these articles in other languages, we use the MediaWiki~API, 
which results in another \num{335809} of articles in all of the 326 languages.
We collect the articles and their complete edit history using the MediaWiki~API in between December 2023 and February 2024.

\subsubsection{Extraction of Source Domains}

Citations may appear in the body of the article text and as references at the bottom of the article, each of which has a standard template.
Each revision that involves a citation contains the author of the revision, position within the document of the citation, text of the citation, and meta information, such as whether it is a DOI, URL, ISBN, ISSN, etc.~or a raw URL.
We standardize these revisions to create a list of ``source edits'' containing the metadata about each insertion, deletion, or edit that involves a source.
We resolve redirects, apply rules to clean and standardize the URLs (including DOI references redirecting to a resource), and merge revisions when the same user edits an article in succession (regardless of the time).
Finally, we extract the domains from the URLs, obtaining \num{376566} unique source domains.




\subsubsection{Dataset Statistics}

The summary statistics about the topic-language datasets (or simply ``datasets'' throughout this paper), for the five topics grouped into low-, mid-, and high-resource languages can be seen in Table~\ref{tab:dataset_stats}. The medians of these values can be found in Appendix~\ref{app:medians}, while more in-depth statistics for each individual language are in Appendix~\ref{app:datasetsizes}.
Recall that we only use
datasets that have passed the threshold of having at least 2 reliable and 2 unreliable perennial sources.

The smallest topic in terms of the number of articles is COVID-19, while the highest numbers of revisions per article are found around Climate Change and COVID-19 in high-resource languages. On the contrary, biology has the largest number of articles, but these are revised on average the fewest number of times in high- and mid-resource languages. High-resource languages have much better coverage of the English perennial sources, with an average of 150 to 260 sources per topic. These values decrease to an average of 15 to 35 sources for low-resource languages, making the training and evaluation more challenging.




\subsection{Feature Definition}

Unlike previous studies measuring Wikipedia edit quality that did not consider certain types of edits and reverts~\cite{trokhymovych2023fair}, we propose to focus on the entire edit history concerning each domain.
Using edit metadata, we define 52 features capturing the popularity of the domain, the ``permanence'' of the domain across edits, and the number and type of users involved in edits adding or removing that domain.
As the features are computed for each domain within each dataset separately (recall, by ``dataset'' we mean a collection of articles on a certain topic, in a certain language) and the datasets are of varying sizes and editorial frequencies, we explore different ways of normalizing these statistics w.r.t.~the first appearance of the domain, age of the dataset, and in terms of time duration or number of revisions. Below, we give a short overview of the features we have used. A more detailed description can be found in Appendix~\ref{appendix_feature_list}.

\subsubsection{Popularity features}
\vspace{-3pt}

\begin{itemize}[leftmargin=*]
\item $N_{\mathrm{articles}}$, $\overline{N_{articles}}$: Number of articles a domain has appeared in (sensitive to the size of the dataset), and its normalized version (by the number of articles in the dataset).

\item $CurrN_{articles}$, $\overline{CurrN_{articles}}$: Number of articles a domain is used in at collection time, and normalized by the total number of articles in the dataset.
\end{itemize}

\subsubsection{Permanence features}

\begin{itemize}[leftmargin=*]
\vspace{-3pt}
\item $\Sigma Perm_d$, $\Sigma Perm_r$: Permanence, how long a domain has been used in an article (not necessarily consecutively), summed across all articles and measured in days (subscript $d$) or by this number of revisions (subscript $r$).

\item $\Sigma CurrPerm_d$, $\Sigma CurrPerm_r$: Same as above, but considering only the articles where the domain is currently used.

\item $\overline{\Sigma Perm_d}$, $\overline{\Sigma Perm_r}$: Sum of all permanences for a domain in all articles, normalized by the sum of the ages of all articles (measured by the number of days or the number of revisions that an article existed).

\item $\overline{\Sigma CurrPerm_d}$, $\overline{\Sigma CurrPerm_r}$: Same as above, but considering only the articles where the domain is currently used.

\item $\langle Perm_d \rangle$, $\langle Perm_r \rangle$: The average permanence of a domain (over all articles where it was used).  

\item $\langle SelfPerm_d\rangle$, $\langle SelfPerm_r\rangle$: Self-permanence: permanence divided by the number of days (or revisions) since the first time the domains were added to the article, then averaged article-wise.  

\item $ \Sigma age_d $, $ \Sigma age_r $: The sum of ages of the domain over all the articles in a dataset, where the age of a domain in an article is the number of days or revisions since a URL from that domain has been first added to the article.

\item $\langle age_d \rangle$, $\langle age_r \rangle$: The average age of a domain over all the articles it appears at least once.
\end{itemize}

\subsubsection{User-based features}

\begin{itemize}[leftmargin=*]
\vspace{-3pt}

\item $U_{add}$, $U_{rem}$: Number of users that have added or removed a domain.

\item $R_{add}$, $R_{rem}$: Same as above, but only counting registered users.

\item $\overline{U_{add}}$, $\overline{U_{rem}}$, $\overline{R_{add}}$, $\overline{R_{rem}}$: The above four features, normalized by the total number of unique users in the dataset.

\item $\langle U_{add} \rangle$, $\langle U_{rem}\rangle$, $\langle R_{add}\rangle$, $\langle R_{rem}\rangle$: The above four features, measured as an average per article.

\item $Ratio(R_{add}/U_{add})$, $Ratio(R_{rem}/U_{rem})$
The proportion of registered vs. all users that ever added or removed a domain.

\item $Proba(R_{add})$, $Proba(R_{rem})$  Probability that, when a domain is added or removed, this is done by the registered users. Contrary to the ratio, here we take into account the number of revisions per user.

\item For all user-based features listed above, we compute a version wherein instead of counting any instance of adding a domain, we only count instances of adding it for the first time on an article (``starting''), and similarly, counting only the instances that the domain was removed for the last time (``ending'').
For example, $Proba(R_{add})$, and $Proba(R_{rem})$ would have a corresponding version $Proba(R_{start})$ and $Proba(R_{end})$ (similarly for the related features).

\end{itemize}

We remark that, to aggregate the various statistics over articles for a domain, we have considered both the average and the median and found the best performance with the average, which we use in the models in the following sections.\footnote{We have also experimented with versions of the above features that use the aggregation over all URLs in each dataset, instead of domains. However, the resulting features were performing very similarly to those aggregated over domains, and we excluded these from the analysis.}

\subsection{Model Training and Evaluation}
\label{sec:model}

Using the above features, we train an XGBoost\footnote{We use Python package XGBoost V1.6.2 \url{https://pypi.org/project/xgboost/1.6.2/}.} model to classify the domains as reliable or unreliable.
To account for class imbalance, we use weights
(e.g., the weight for the positive class is the fraction of negative/positive).
To limit overfitting, we set the maximum depth of the classifier to 1 
(we have tested the classifier with depths up to 5, but found no increase in leave-one-out (LOO) validation performance, 
and the learning rate to $\eta=0.1$ (default is 0.3).

When evaluating in the ``native'' condition, wherein the training and target data are from the same dataset, we employ LOO cross-validation.
In the ``cross-language adaptation'' condition, wherein the model is trained on one language and tested on another (keeping the topic the same), we simply compute the evaluation metrics on the test language.
Similarly, for cross-topic adaptation, we keep the language consistent and test on a target topic. 
In the ``mixed'' condition, wherein the model is trained on a set of languages and tested on one of these languages, we again employ LOO cross-validation where a domain of test language is left out at a time. 

To compute a confidence interval for a metric, we employ bootstrapping wherein we resample with replacement the output of the classifier and recompute the metrics $n=100$ times.
The main metric we compute is the F1 macro score, which is the unweighted average of the F1 scores of the negative and positive classes.

\section{Results}
\label{sec:results}


\begin{table*}[ht!]
\caption{Leave-one-out validation F1 macro, and precision and recall metrics for both classes for the model trained and tested in the English Wikipedia on each topic, and for all topics combined (standard deviations specified with $\pm$). Class balance in terms of the number of reliable (rel) and unreliable (unr) domains. }
\label{tab:loo_native}
\centering
\footnotesize
\begin{tabular}{l|ccccc|cc}
\toprule
Topic          & F1  Macro               & Precision (rel)     & Recall (rel)       & Precision (unr)     & Recall (unr)    & \# rel & \# unr    \\ \midrule
Climate Change & $0.81 \pm 0.02$    & $0.80 \pm 0.03$     & $0.80 \pm 0.04$    & $0.83 \pm 0.04$     & $0.83 \pm 0.03$   & $129$ &  152\\
COVID-19       & $0.88 \pm 0.02$    & $0.87 \pm 0.03$     & $0.91 \pm 0.03$    & $0.89 \pm 0.03$     & $0.85 \pm 0.03$  & $136$ &  125 \\
Biology        & $0.80 \pm 0.03$    & $0.81 \pm 0.03$     & $0.81 \pm 0.03$    & $0.80 \pm 0.03$     & $0.80 \pm 0.03$  & $124$ &  120 \\
History        & $0.75 \pm 0.03$    & $0.70 \pm 0.04$     & $0.83 \pm 0.04$    & $0.82 \pm 0.03$     & $0.69 \pm 0.05$  & $120$ &   139\\
Media          & $0.83 \pm 0.02$    & $0.81 \pm 0.03$     & $0.82 \pm 0.03$    & $0.85 \pm 0.03$     & $0.84 \pm 0.03$  & $146$ &  176 \\ \midrule
All topics      & $0.83 \pm 0.02$    & $0.76 \pm 0.04$     & $0.85 \pm 0.03$    & $0.89 \pm 0.02$     & $0.82 \pm 0.03$  & $159$ &  234 \\ \bottomrule
\end{tabular}

\end{table*}

\begin{figure*}[ht!]
  \centering
    \subfloat[Feature Importance\label{fig:RQ2_beeswarm}]{\includegraphics[width=0.60\linewidth]{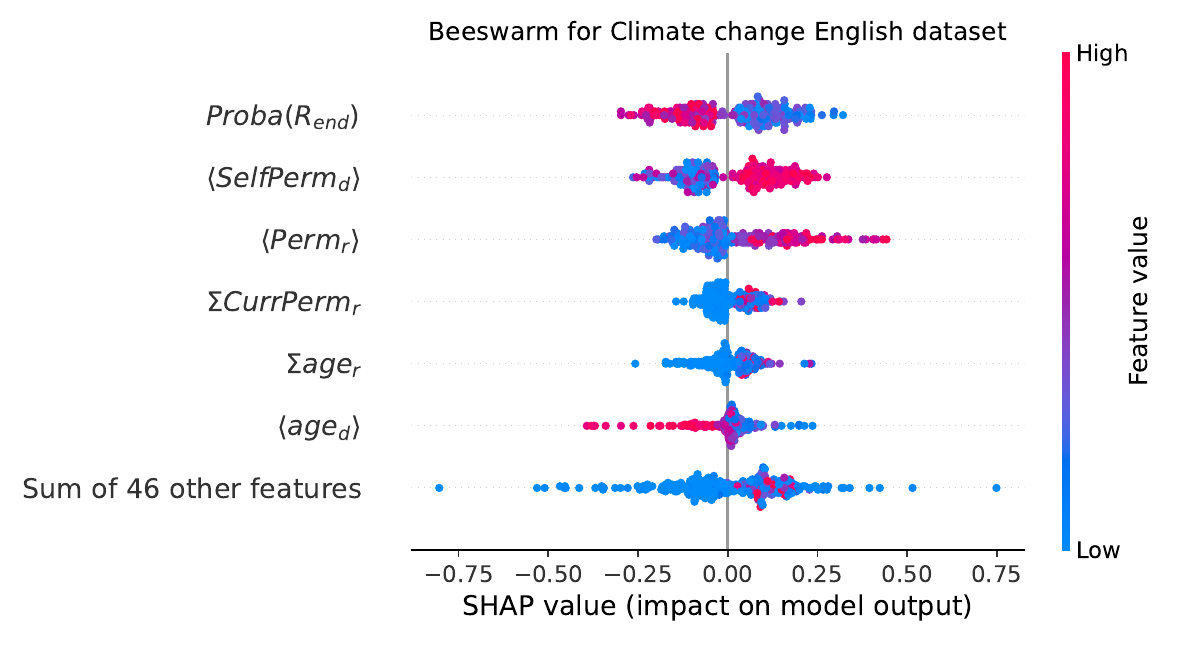}}\\\vspace{0.5cm}
    \subfloat[twitter.com\label{fig:RQ1_twitter}]
    {\includegraphics[width=0.49\linewidth]{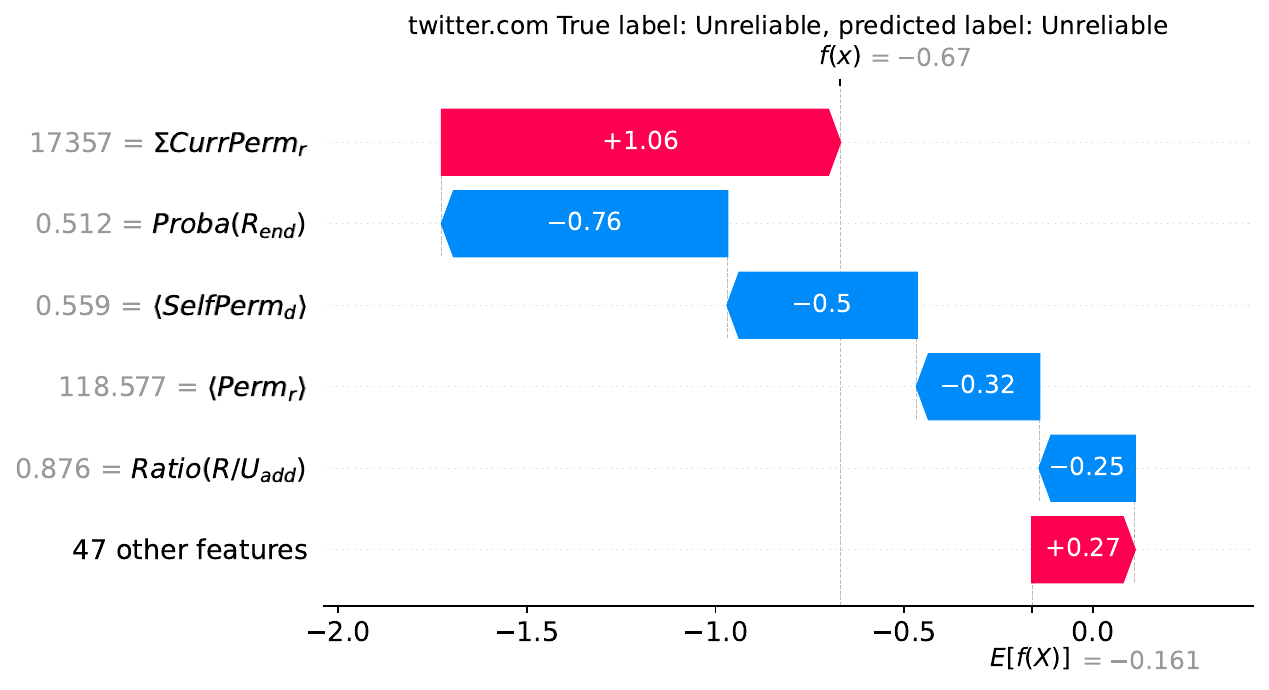}}
    \subfloat[researchgate.net\label{fig:RQ1_Researchgate}]{\includegraphics[width=0.49\linewidth]{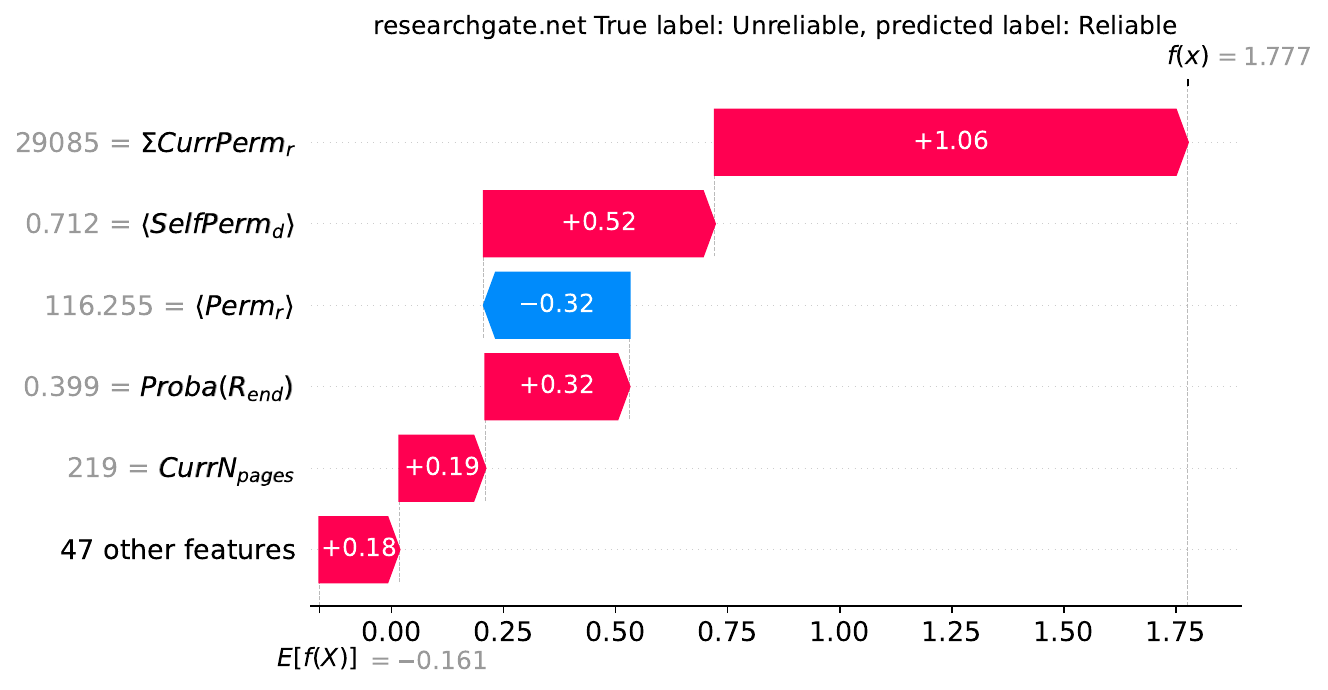}}
  \caption{Beeswarm analysis of an XGBoost model trained on Climate Change English dataset (class encoding is 1: reliable and 0: unreliable) and two example classification explanations (score shown is the log-odds ratio).  }
    \Description[<short description>]{<long description>} 
  \label{fig:SHAP}
\end{figure*}

\subsection{RQ1: Case study: Climate Change in English}


We begin by studying the behavior of our model for the Wikipedia articles in English around the Climate Change topic.
The model performance is shown in the first row of Table~\ref{tab:loo_native}, which lists the F1 Macro, precision and recall for each class.
We find that, out of the unreliable domains, the classifier correctly identifies as unreliable 83\% of them (recall), and out of the domains the classifier guessed as unreliable, 83\% are truly unreliable (precision). Both precision and recall for the reliable class are slightly worse at 80\%.
Note that the classes are fairly balanced in this setting, with 46\% of the dataset having the reliable label.

Figure~\ref{fig:RQ2_beeswarm} shows the ``bee swarm'' plot of SHAP (SHapley Additive exPlanations) values that illustrate the importance of each feature in the classifier~\cite{NIPS2017_7062}.
The top 6 most predictive features (and an aggregation of the rest) are shown.
The SHAP values for the top feature $Proba( R_{end} )$ -- the probability that, when a domain is removed, it is done by a registered editor -- reveal that high values of this feature are associated with a negative impact on the domain reliability prediction.
In other words, when a registered editor removes a domain, it is more likely to be unreliable.
The opposite can be seen for the $\langle SelfPerm_d \rangle$ --average proportion of days the domain remains cited since it was first added-- and $\langle Perm_r \rangle$ --the average number of revisions that the domain remains on the article--: higher values of these features are associated with more reliable sources.
$\Sigma CurrPerm_r$ --the aggregated number of revisions through which the domain has existed across all articles-- also correlates positively with the reliable category, as does as well the sum of the domain ages across articles, $\Sigma age_r$.
Finally, ``younger'' domains -- those that have been added to an article more recently, $\langle age_d \rangle $ -- are more likely to be reliable.
This may be a sign of the increased focus of the platform on the quality of sources~\cite{Baigutanova_2023}.

In Figures~\ref{fig:RQ1_twitter} and~\ref{fig:RQ1_Researchgate} we illustrate the performance of the model for two domains of special interest: \emph{twitter.com}, correctly identified as unreliable, and \emph{researchgate.net}, incorrectly identified as reliable.
In the case of \emph{twitter.com}, despite being added/present frequently on articles (high $\Sigma CurrPerm_r$), it is soon removed by registered editors ($Proba( R_{end} )$ and also SelfPermanence, $\langle SelfPerm_d \rangle $ is low).
For \emph{researchgate.net}, on the other hand,  model and ground truth disagree: most of the permanence features indicate that it is used as a reliable source, while the domain is considered unreliable in the perennial source list because of being a self-publishing platform lacking editorial oversight or peer review. However, as our data points out,  Wikipedia editors may still use it pragmatically, for example, to link to freely accessible versions of peer-reviewed papers that are otherwise behind paywalls. This example underscores the potential of our method to reveal discrepancies between Wikipedia’s formal sourcing guidelines and how editors apply them in practice, highlighting edge cases that may warrant further discussion or refinement in policy.



\subsection{RQ2: Topic and Language Generalizability}

We continue by evaluating the performance of our language-agnostic modeling approach in other topical and linguistic settings.
Table~\ref{tab:loo_native} shows the F1 macro and class-specific precision and recall for these topics, in the English Wikipedia.
We find that, overall, the performance is consistent, with the lowest performance observed for the History dataset,  with a recall of 0.69 for the unreliable class.
The best performance is achieved on the COVID-19 dataset, possibly due to the ample data on the edits and the more recent nature of the corresponding articles. We find that the amount of edited data is important for the classifier's performance, as we show later in this section. 
Finally, we combine all English-language datasets into one and perform training and LOO validation; the performance of this model is shown in Table~\ref{tab:loo_native} under ``all topics''.
Such a combined model performs on par with other topical models, with especially high precision for the unreliable class (0.89), suggesting that combining knowledge about different topics does not overall degrade the performance of the classifier.

\begin{figure}[t]
  \centering
  \includegraphics[width=0.8\linewidth]{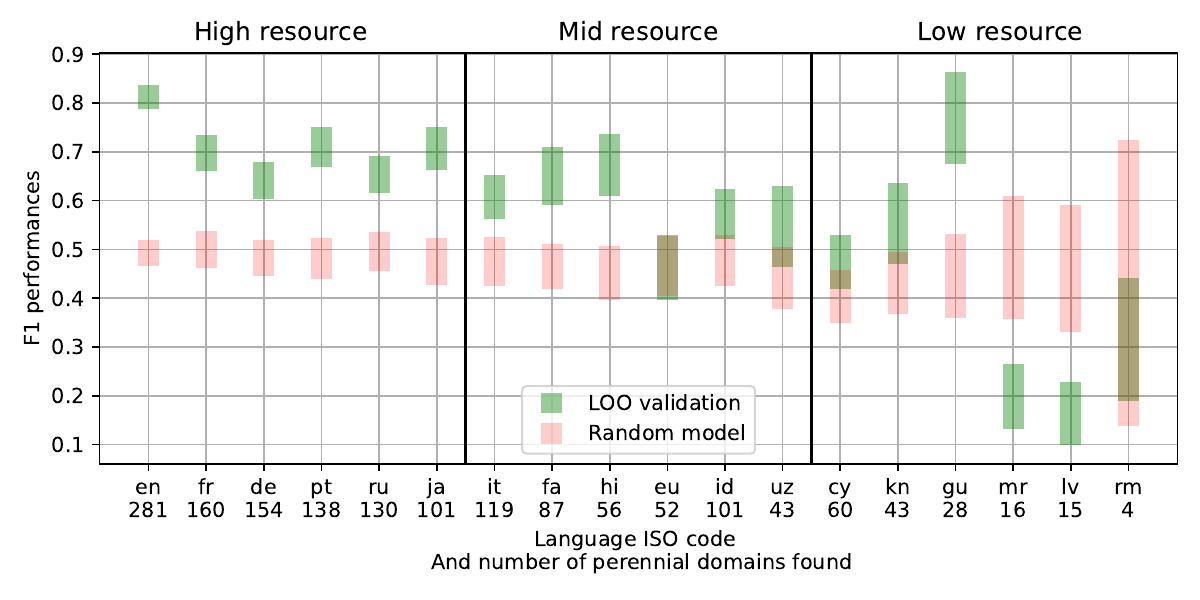}
  \caption{Performances (mean $\pm$ stdv) of models trained and validated in languages other than English on the Climate Change datasets. The three batches of languages, separated by a vertical black line, show results on a sample of high (left), mid (center), and low (right) resource languages.}
  \Description[<short description>]{<long description>} 
  \label{fig:RQ3_barplot}
\end{figure}

Next, we explore the applicability of the proposed model to other languages.
In these experiments, we first keep the topic constant -- Climate Change -- while training and evaluating models in different languages.
Figure~\ref{fig:RQ3_barplot} shows the LOO F1 macro scores for select languages form the high-, mid-, and low-resource groups, in comparison to the performance of a random baseline model.
Specifically, we show the average performance $\pm$ one standard deviation as shaded areas.
To estimate the performance of a random model, we assign a random binary classification for the domains and perform bootstrapping ($n=100$) similar to the evaluation of all other models, as discussed in Section~\ref{sec:model}.
The performance is highest for high-resource languages and degrades for mid- and low-resource ones.
The standard deviations of the scores also increase with lower resource languages (both for our model and the baselines), pointing to an increased lack of information for computing the score.

\begin{figure}[t!]
  \centering
  \includegraphics[width=0.8\linewidth]{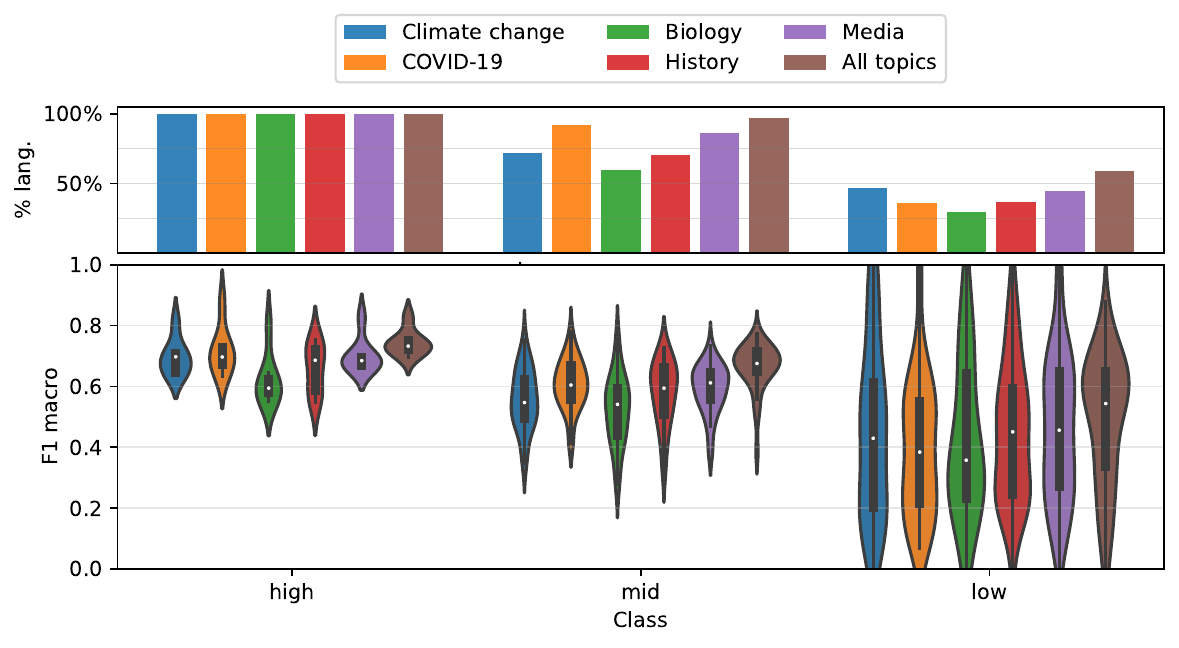}
  \caption{Native model performance per topic for different language types. Top panel: \% of models that perform better than a random classifier. Bottom panel: Violin plots of corresponding distributions of F1 macro performances.}
  \Description[<short description>]{<long description>} 
  \label{fig:RQ2_MW}
\end{figure}

Additionally, we show aggregate statistics of the model performance on the languages in each category in Figure~\ref{fig:RQ2_MW}.
In particular, we are interested in whether our model outperforms the random classifier in terms of F1 macro score (top panel of the figure): we perform this comparison using Mann-Whitney scores and adjust the significance level of 0.05 using the Bonferroni correction for multiple comparisons.
For instance, in the Climate Change topic, the model outperforms the random baseline in 100\% of the high-resource languages, but only in 44\% of low-resource languages (in an additional 12\% the difference is not statistically significant).
The performance on low-resource languages tends to vary widely for different topics.
The brown bars show the models for which articles for all topics within a language are used for training and testing. These models perform markedly better than topic-specific ones, echoing our findings for the English language.

\begin{figure}[ht!]
  \centering
  \includegraphics[width=0.60\linewidth]{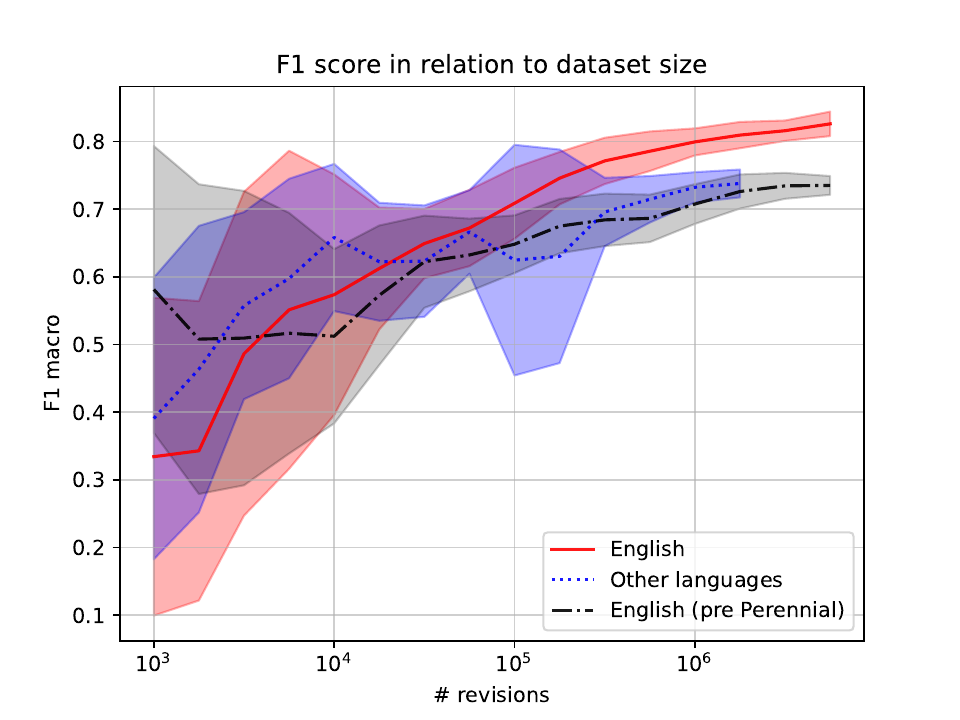}
  \caption{Average model F1 macro (line) and standard dev. (shaded area) versus the size of the training dataset in terms of the number of revisions. In English (red), all topics are combined, and the data is sampled at regular intervals. The same experiment is repeated, considering only revisions before the implementation of the perennial sources on 2018-07-01 (grey). The performance of models in all other languages (also all-topic) is shown in blue, without sub-sampling.}
  \label{fig:increasing_experiment}
    \Description[<short description>]{<long description>} 
\end{figure}

\subsubsection*{Effect of dataset size and introduction of perennial sources on model performance}

The worse performance of our model in the lower-resource languages may be mainly (among others) due to two reasons: the datasets available for training the model are too small, or the behavior of the users of these languages towards domains is intrinsically difficult to model (their treatment of reliable and unreliable domains is difficult to distinguish).
We design two experiments to address these two reasons, starting with the model's sensitivity to the amount of training data.

The red line in Figure~\ref{fig:increasing_experiment} shows the performance of the model trained on different dataset sizes. In particular, we consider the English (all-topic) dataset and sample the data at log-regular intervals (starting at $10^3$, then at $10^{3.25}$ etc., until $10^{6.75}$), resulting in an increasing number of revisions available for training. We repeat this exercise 10 times (mean $\pm$ stdv. shown in the Figure).
For comparison, we depict the performance of the models trained on the all-topic datasets in all other languages (shown in blue). Note that in these cases, there is no sampling, and for every language, all available data is used.
We find that the performance of the English-language model is comparable to that of the other languages when there is little data available (up until about $10^5$ revisions) and is about 10\% higher than the other languages when almost all data is used for its training.
At least in the English-language case, we find that for stable model performance
around 10,000 revisions are necessary.
At the higher range of data, we find that the classifier's performance has not leveled off, suggesting that adding more articles (possibly other topics) may improve the performance further.

Turning to editing behavior, the introduction of the perennial sources list provides a ``natural experiment'' to measure a possible editing behavior change in response to these additional guidelines.
We only examine the impact of this change in the English-language datasets, as it is not
clear whether editors from other language editions of Wikipedia used the English perennial sources list.
Thus, we created a version of our English-language dataset using only revisions up to 2018-06-30 (i.e., before these guidelines were introduced).
To control for the dataset size (which, as shown above, may affect model performance), we again perform sub-sampling similar to what we did with the full English-language dataset.
The black dash-dotted line and the grey area in Figure~\ref{fig:increasing_experiment} show the model performance based on this limited data.
We observe its performance to be notably lower than with more recent data (red line and area), indicating that the editing behavior before the introduction of the perennial sources list was less predictive of reliable/unreliable domain classes than afterwards.
In this case, the performance is similar to that of other languages, showing that editor behavior allows extracting meaningful and predictive signals for source reliability even when the editors did not have an official guideline for source reliability.
Furthermore, these findings support previous work, which shows that the quality of the Wikipedia references improved after the introduction of the perennial sources~\cite{Baigutanova_2023}.

\subsection{RQ3: Model Adaptation across Topics and Languages}

\begin{figure}[t]
  \centering
  \includegraphics[width=0.8\linewidth]{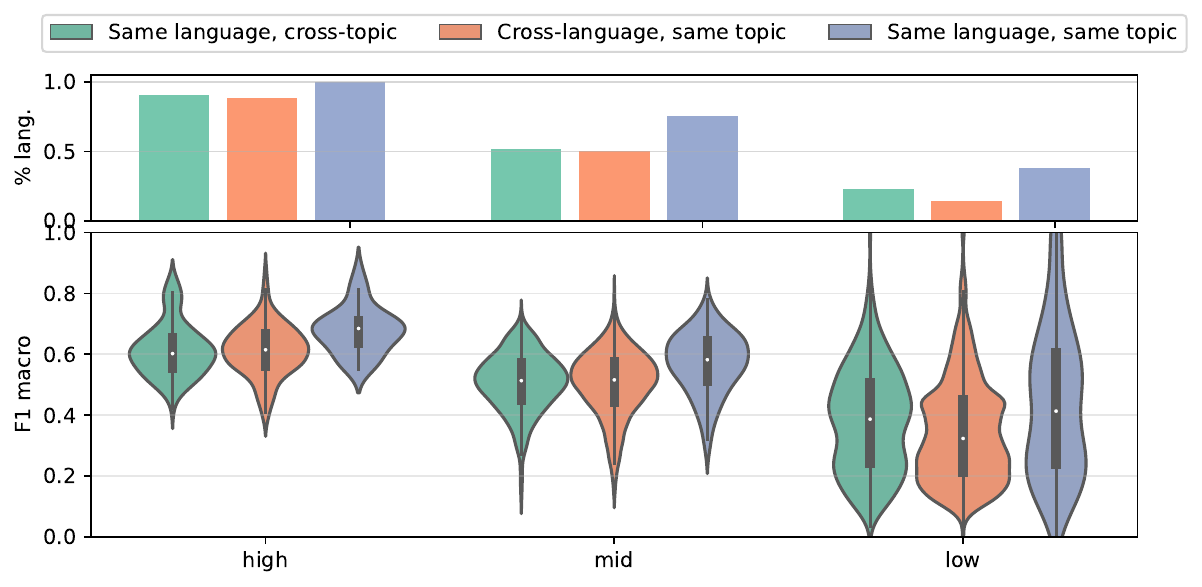}
  \caption{Model performance in two adaptation scenarios: cross-language setting (trained on a language, tested on a different language from the same class, in the same topic, in red) and a cross-topic setting (trained on a topic, tested on a different topic, in the same language, in green). Results aggregated by language resourcefulness and compared to same-topic and language native models (in blue). }
  \label{fig:RQ4_crossdataset}
    \Description[<short description>]{<long description>} 
\end{figure}

The multilingual nature of Wikipedia presents an opportunity to supplement information available in lower-resource languages with that in the larger ones.
Considering our datasets, two directions of adaptation are possible: across languages and across topics.
We consider both scenarios in Figure~\ref{fig:RQ4_crossdataset}, which shows the performance of the models adapted in these two ways: in a cross-language setting (trained on a language and afterwards tested on a different language from the same resourcefulness class, in the same topic) and a cross-topic setting (trained on a topic, tested on a different topic, in the same language).
We find that, in both cases, the performance of the adapted models decreases compared to the native performance.
The decrease in performance is about the same in cross-language and cross-topic settings.
This suggests that the editor's behavior around references may be non-identical 
in different topics, as well as languages. 
However, again, we see the importance of resourcefulness on the performance: the models adapted across high-resource languages perform better than those across mid and low-resource ones.

\begin{figure}[t]
  \centering
  \includegraphics[width=0.80\linewidth]{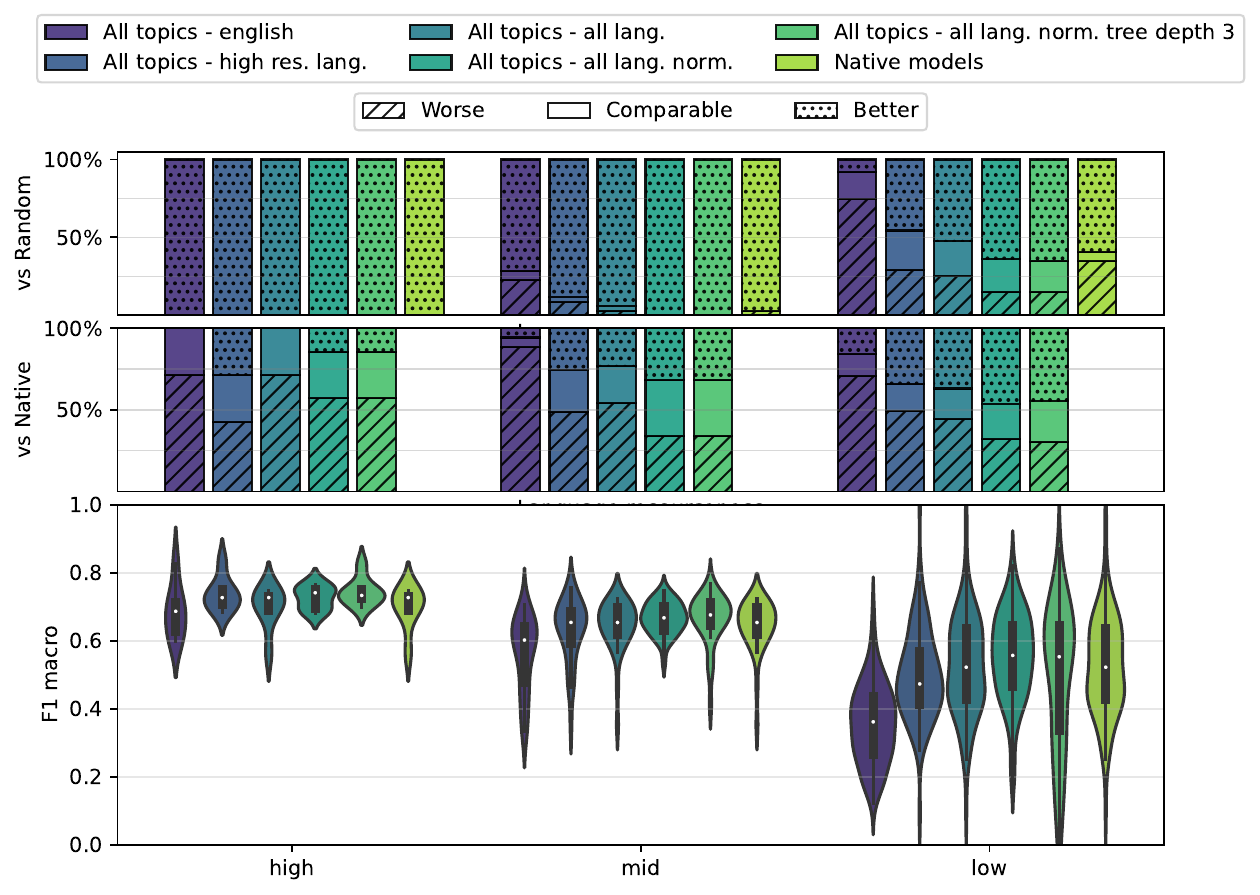}
  \caption{Combined Model performances for different training datasets and strategies. Top panel: \% of models that perform better, same as, or worse than a random classifier; Center: same, but compared to the native model; Bottom: violin plots of the corresponding distributions of the F1 macro performances. }
  \label{fig:RQ3_experiments}
  \Description[<short description>]{<long description>} 
\end{figure}

Finally, we consider all topics together and analyze the performance of our model when trained with different training strategies and tested on each language.
In particular, we test whether models trained on higher-resource data would be helpful to those in lower-resource settings.
In Figure~\ref{fig:RQ3_experiments} we aggregate the corresponding results by languages of similar resourcefulness: the top panel shows how for many languages the different models perform significantly better, same as, or worse than a random model (according to Mann-Whitney tests with Bonferroni correction), the center panel shows the same, but compared to the native model, and the bottom panel shows the distribution of F1 macro scores.
We find that F1 scores are generally lower when training a model on the English dataset and applying it to low-resource languages, confirming that cross-language adaptation can be problematic, even if a high-resource language such as English is used.
However, when training on all high-resource languages or all languages together, the performance for low-resource languages significantly increases.
This suggests that including a plurality of languages in training data may capture a variety of user behaviors that are transferable to lower-resource languages.

Nonetheless, each language-specific dataset has its own biases connected to the amount of data available, temporal peculiarities of the articles and their edits, and the volume of editing activity.
To alleviate potential difficulties in adapting feature values between languages, we normalize each dataset using quantile normalization (converting feature values to ranks).
This approach further improves the performance of the all-language model adaptation, even on average
performing better than the native models. In more detail, the model outperforms a random classifier in more than 65\% of low-resource settings, and even outperforms native models in 47\% of these cases, while being on par with it in 21\%. Notably, models trained solely on English fail to generalize well to low-resource languages, performing worse than random in over 70\% of cases—highlighting the distinct editorial behaviors across linguistic communities.

Finally, since the size of the dataset increases with the combination of topics, we train the classifier with the maximum tree depth of 3 instead of 1, but do not find a substantial increase in performance.

\subsection{In-Depth Examination of Source Usage}
\label{sec:qualitative}

The above quantitative evaluation of the proposed model abstracts the rich insight that its output provides.
When examining the domains that attain the highest reliability score by the classifier, we find interesting signals of the editorial treatments of different sources, in each topical sphere (here, we focus on English data).
For instance, among the domains predicted (incorrectly) as reliable in the Climate Change domain is YouTube (which appeared in 644 articles in that dataset)
with a positive (reliable) class probability of $P_r=0.64$,
as well as Internet Movie Database (IMDB) (appearing in 56 articles, $P_r=0.56$). This points to the importance of popular or mainstream culture in this topic.
When considering the History dataset, we find the classifier incorrectly labels as reliable domains pointing to sources that the perennial sources label as having partisan bias, such as the Jewish Virtual Library ($P_r=0.81$), Sixth Tone ($P_r=0.74$), and Daily Sabah ($P_r=0.69$).

On the other hand, one can use the output of this classifier to find sources that different communities of editors consider inappropriate for their topical domain.
For instance, in the Biology dataset, Science-Based Medicine, which is considered generally reliable and has ``a credible editorial board'' according to the perennial source list, is treated by the editors more similarly to unreliable domains ($P_r=0.20$).
Similarly, the Huffington Post, rated as ``fairly reliable for factual reporting on non-political topics'', scores low in the COVID dataset ($P_r=0.07$), possibly due to the politically charged and serious nature of the rhetoric around the pandemic.
As editorial policies of sources, as well as perceptions and behaviors of the Wikipedia editors, change, the metrics developed in this study may provide a guide for the improvement and updating of the perennial sources and potential adjustment of the labels in specific domains of expertise.

\subsection{Alternative Indicators of Source Reliability}

One limiting factor of our study was the availability of labels for the quality of sources.
We consider an alternative: Media Bias / Fact Check (MBFC), a third-party website that provides ratings of ``the bias, factual accuracy, and credibility of media sources''\footnote{\url{https://mediabiasfactcheck.com/about/}}, which has been widely used in misinformation research and by other indexes~\cite{iffy,chowdhury2020joint,ye2019mediarank}.
We use this credibility metric 
and consider \emph{very high} and \emph{high} as the positive class, and \emph{low} and \emph{very low} as negative.
With the same features and model as above and LOO cross-validation, we gauge its performance on MBFC.
Although the class-average F1 macro is 0.61, the classifier performs especially poorly on the negative class, with a precision of 0.32 and recall of 0.55, suggesting that the features do not capture this metric.
Indeed, there may be several reasons why such resources are not suitable for the analysis of Wikipedia references.
First, there is little overlap between the sources noted in MBFC and in the English Wikipedia perennial sources list (out of 1156 perennial sources, 272 are in MBFC). 
Second, the MBFC credibility metric is weakly aligned with the perennial source reliability one.
Third, MBFC itself is listed as \emph{generally unreliable} by the English Wikipedia perennial sources list.

In conclusion, MBFC and similar resources are valuable for assessing source reliability in general research contexts. However, the above observations suggest that they are less suitable for evaluating the reliability of sources in the specific context of Wikipedia articles, especially if used as ground truth for supervised classification methods like those presented in this work.




\section{Discussion}
\label{sec:discussion}

In this study, we have proposed and assessed editing behavior-based features for modeling the reliability of web domains used as references on Wikipedia.
 We show that these models can be a part of a language-agnostic toolkit applicable both to high- and low-resource languages.
 As such, this work extends previous efforts to use the edit history to measure the controversiality of Wikipedia articles~\cite{borra2015contropedia}.
Editing behavior is not the only source of useful information
for language-agnostic modeling.
Links between articles (already used for topical clustering of Wikipedia articles~\cite{johnson2021language}) may provide semantically meaningful context for references,
whereas users' reading sessions (used for entity linking~\cite{gerlach2021multilingual}) may give controversiality signals from the perspective of the audience.
Future work could also take a page from the social media literature and build editor-resource networks~\cite{shu2019studying} to capture structural peculiarities of topics, editors, and resources associated with low-quality content.
Additionally, automated feature discovery methods, such as self-supervised or representation learning on structured edit data, could help uncover latent behavioral patterns not easily captured by manually designed features. However, this would come at the cost of losing interpretability, a key value of our current approach.

Apart from finding that the proposed features capture editing behavior within a dataset spanning one topic and language, we also observe that they retain useful information during the model adaptation across languages.
Although English has been the \emph{lingua franca} in NLP, having the best-developed tools and resources, we show that adaptation from the English-language dataset to other languages is largely inferior to using a model that combines information from many languages.
This improvement is especially noticeable in performance on low-resource languages, where native models often perform no better than random baselines, due to insufficient revision and citation activity for effective learning. However, when training on data pooled from all languages and applying quantile normalization to account for resource differences, the model significantly outperforms random baselines and matches or outperforms native models in most low-resource cases. This strategy highlights a practical and scalable pathway to empower under-resourced Wikipedia communities, even in the absence of extensive local data or tooling.

These findings are aligned with limitations identified by~\citet{das2024language}, who suggest that, although language-agnostic features are used, limiting data selection to one language may limit the breadth of the captured behaviors and their applicability to new settings.
As multilingual embedding models have gained popularity~\cite{zhang2024m3exam}, it remains to be seen to what extent the latest developments in NLP will benefit the low-resource languages~\cite{zhang2023don}, and what the impact will be on Wikipedia specifically~\cite{johnson2024wikimedia}. In this context, the need to monitor possible changes in the editorial standards and citation practices across languages becomes especially important. 
Community efforts could prioritize on developing robust cross-lingual transfer strategies, rather than relying solely on local data that may be too sparse to train reliable predictors




As ground truth, we have used the list of perennial sources~\cite{perennial_sources} from the English Wikipedia, which is part of Wikipedia's efforts to provide clear editorial instructions to its community.
We compare this resource with another popular source of reliability measure, the MBFC~\cite{mbfc}, 
and find significant differences between them, pointing to the necessity of
Wikipedia-specific models like ours.
Previous work comparing the perennial sources lists in different languages showed disagreement on whether some domains are indeed reliable~\cite{baigutanova2023comparative}.
Therefore, the system we propose could be used to expand lists of perennial sources and lists derived from them\footnote{\url{https://science.feedback.org/consensus-credibility-scores-comprehensive-dataset-web-domains-credibility/}}, or even to assess cultural biases in Wikipedia sources~\cite{zheng2023gender,yang2024polarization}.

\subsection*{Limitations}

The present study has multiple limitations.
First, using language-agnostic approaches may be seen as a limitation, as it does not take advantage of the predictive power of language-dependent features. However, despite the significant advances with multilingual NLP resources such as mBERT~\cite{devlin2018bert}, the motivation for our model was precisely to be ready for use in the 300+ language editions of Wikipedia, especially those that are less privileged and under-resourced.

Second, while the dataset contains thousands of articles and cumulatively millions of revisions, it spans only a fraction of Wikipedia.
Despite choosing topical foci such that they have different reference quality and time scales, they may not capture the entire diversity of subject matter of this massive collaborative effort.
For instance, differentiating topics by whether the topic is currently controversial may reveal shifts in editorial handling of sources. Additionally, future versions of our modeling approach could incorporate newly generated data by implementing re-training strategies using updated activity data, including changes in the classification of web domain reliability scores from the perennial source list.

Third, our approach may inherently favor generalist sources over topic-specific ones. Because several behavioral features (e.g., the number of articles or revisions in which a source domain appears) capture overall prevalence within Wikipedia, broadly used domains tend to accumulate stronger permanence and visibility signals. Conversely, specialized domains, despite being authoritative within their fields, may appear in fewer contexts and thus receive lower reliability estimates. This limitation suggests a potential bias toward sources covering a wider topical range. Its presence could be investigated in future research and mitigated by normalizing the corresponding features by topical diversity.

Fourth, the selection of the English perennial sources list as the gold standard disadvantaged other languages.
Unfortunately, only a few other languages have a similar resource, and they are significantly less extensive and conclusive~\cite{baigutanova2023comparative}.
Given the global role played historically by the English edition of Wikipedia~\cite{hale2014multilinguals}, our selection provided the best choice to reach the maximum potential coverage across other language editions.
However, the adaptation of the domain list from English may not coincide with possibly different cultural or local interpretations of quality by the editors from other language communities.
In such cases, it may be preferable to apply NLP transfer methods~\cite{das2024effective} that leverage higher-resource language editions to support lower-resource ones, particularly when they are comparable in the domain of source reliability. Our study, which employs language-agnostic features, can serve as a foundation for identifying appropriate language pairs for transfer in future work.

Fifth, our model relies on a fixed ground truth, consistent with prior work (e.g., \cite{Baigutanova_2023,baigutanova2023comparative}). However, perennial source lists evolve over time as editorial consensus changes and new sources are evaluated. While we treat it as static for this study, future work should incorporate time-aware labels and explore retraining strategies to adapt to such changes and maintain alignment with current community standards.

Finally, we show that the method is sensitive to the amount of data available for training (and testing), possibly resulting in insignificant results in some low-resource language settings. Further research will be needed to extend the amount of available data in these contexts.

\subsection*{Ethical Considerations}

This study, and online content moderation in general, have important ethical aspects that should be kept in mind by both the platforms and the misinformation researchers.

First, this study has a direct impact on the access equity to encyclopedic resources in low-resource languages.
Currently, the lack of resources, such as the perennial source list for these versions of Wikipedia, limits the extent to which our model is able to reflect the standards of that linguistic community.
The domain scores provided by the models adapted from other (higher resource) languages that we propose here could be a starting point for bootstrapping the editorial discussion of the quality of sources used in low-resource languages.

Second, the process of determining the quality of sources bears scrutiny.
In the news domain, the quality of information is often determined by professional fact-checkers.
However, on Wikipedia, the editors from the community define the status of the sources via a deliberation process, guided by the community standards.
Any potential biases in the judgment of these editors or failings in the deliberative process would be reflected in the quality of these labels (for instance, according to~\citet{kharazian2023governance}, some Wikipedia editions were dominated for a time by a ``small group of administrators who introduced far-right bias and outright disinformation'').
Generally, in 2014~\citet{shaw2014laboratories} concluded that the Wikipedia contributors self-organize into oligarchic structures, though in 2021~\citet{rijshouwer2023wikipedia} found ``strong counter-tendencies'' to editorial power concentration.
Our proposed method may reveal community behaviors that are not necessarily explicitly stated, but which shape the platform's use of outside sources, potentially aiding the community to self-monitor the use of sources.

We believe that the code and resources we make available to the community\footnote{\url{https://github.com/JacopoDignazi/Wiscom}} (compiled in accordance with FAIR (Findable, Accessible, Interoperable, and Reusable  principles~\cite{wilkinson2016fair}) will be useful to assist future research efforts and to further evaluate and extend the features and modeling we proposed.

\begin{acks}
This work is partially supported by a grant from the Credibility Coalition, a project of Meedan and Hacks/Hackers. Authors also acknowledge support from the Lagrange Project of the Institute for Scientific Interchange Foundation (ISI Foundation), funded by Fondazione Cassa di Risparmio di Torino (Fondazione CRT). Earlier versions of this paper benefited from valuable feedback from Diego Sáez Trumper, Francisco Navas, Aitolkyn Baigutanova, and attendees of Wiki Workshop~2024.
\end{acks}

\bibliographystyle{ACM-Reference-Format}
\bibliography{references}


\begin{thebibliography}{59}


\ifx \showCODEN    \undefined \def \showCODEN     #1{\unskip}     \fi
\ifx \showDOI      \undefined \def \showDOI       #1{#1}\fi
\ifx \showISBNx    \undefined \def \showISBNx     #1{\unskip}     \fi
\ifx \showISBNxiii \undefined \def \showISBNxiii  #1{\unskip}     \fi
\ifx \showISSN     \undefined \def \showISSN      #1{\unskip}     \fi
\ifx \showLCCN     \undefined \def \showLCCN      #1{\unskip}     \fi
\ifx \shownote     \undefined \def \shownote      #1{#1}          \fi
\ifx \showarticletitle \undefined \def \showarticletitle #1{#1}   \fi
\ifx \showURL      \undefined \def \showURL       {\relax}        \fi
\providecommand\bibfield[2]{#2}
\providecommand\bibinfo[2]{#2}
\providecommand\natexlab[1]{#1}
\providecommand\showeprint[2][]{arXiv:#2}

\bibitem[\protect\citeauthoryear{Arag{\'{o}}n and S{\'{a}}ez{-}Trumper}{Arag{\'{o}}n and S{\'{a}}ez{-}Trumper}{2021}]%
        {aragon2021preliminary}
\bibfield{author}{\bibinfo{person}{Pablo Arag{\'{o}}n} {and} \bibinfo{person}{Diego S{\'{a}}ez{-}Trumper}.} \bibinfo{year}{2021}\natexlab{}.
\newblock \showarticletitle{A preliminary approach to knowledge integrity risk assessment in Wikipedia projects}.
\newblock \bibinfo{journal}{\emph{CoRR}}  \bibinfo{volume}{abs/2106.15940} (\bibinfo{year}{2021}), \bibinfo{pages}{1--4}.
\newblock
\showeprint[arXiv]{2106.15940}
\urldef\tempurl%
\url{https://arxiv.org/abs/2106.15940}
\showURL{%
\tempurl}


\bibitem[\protect\citeauthoryear{Arora, West, and Gerlach}{Arora et~al\mbox{.}}{2024}]%
        {arora2024orphan}
\bibfield{author}{\bibinfo{person}{Akhil Arora}, \bibinfo{person}{Robert West}, {and} \bibinfo{person}{Martin Gerlach}.} \bibinfo{year}{2024}\natexlab{}.
\newblock \showarticletitle{Orphan articles: The dark matter of wikipedia}. In \bibinfo{booktitle}{\emph{Proceedings of the International AAAI Conference on Web and Social Media}}, Vol.~\bibinfo{volume}{18}. \bibinfo{publisher}{AAAI Press}, \bibinfo{address}{Washington, DC, USA}, \bibinfo{pages}{100--112}.
\newblock


\bibitem[\protect\citeauthoryear{Ashkinaze, Gilbert, and Budak}{Ashkinaze et~al\mbox{.}}{2024}]%
        {ashkinaze2024dynamics}
\bibfield{author}{\bibinfo{person}{Joshua Ashkinaze}, \bibinfo{person}{Eric Gilbert}, {and} \bibinfo{person}{Ceren Budak}.} \bibinfo{year}{2024}\natexlab{}.
\newblock \showarticletitle{The Dynamics of (Not) Unfollowing Misinformation Spreaders}. In \bibinfo{booktitle}{\emph{Proceedings of the ACM on Web Conference 2024}}. \bibinfo{publisher}{Association for Computing Machinery}, \bibinfo{address}{New York, NY, USA}, \bibinfo{pages}{1115--1125}.
\newblock


\bibitem[\protect\citeauthoryear{Baigutanova, Myung, Saez-Trumper, Chou, Redi, Jung, and Cha}{Baigutanova et~al\mbox{.}}{2023a}]%
        {Baigutanova_2023}
\bibfield{author}{\bibinfo{person}{Aitolkyn Baigutanova}, \bibinfo{person}{Jaehyeon Myung}, \bibinfo{person}{Diego Saez-Trumper}, \bibinfo{person}{Ai-Jou Chou}, \bibinfo{person}{Miriam Redi}, \bibinfo{person}{Changwook Jung}, {and} \bibinfo{person}{Meeyoung Cha}.} \bibinfo{year}{2023}\natexlab{a}.
\newblock \showarticletitle{Longitudinal Assessment of Reference Quality on Wikipedia}. In \bibinfo{booktitle}{\emph{Proceedings of the ACM Web Conference 2023}} \emph{(\bibinfo{series}{WWW ’23})}. \bibinfo{publisher}{ACM}, \bibinfo{address}{New York, NY, USA}, \bibinfo{pages}{2831–2839}.
\newblock
\urldef\tempurl%
\url{https://doi.org/10.1145/3543507.3583218}
\showDOI{\tempurl}


\bibitem[\protect\citeauthoryear{Baigutanova, Saez-Trumper, Redi, Cha, and Arag\'{o}n}{Baigutanova et~al\mbox{.}}{2023b}]%
        {baigutanova2023comparative}
\bibfield{author}{\bibinfo{person}{Aitolkyn Baigutanova}, \bibinfo{person}{Diego Saez-Trumper}, \bibinfo{person}{Miriam Redi}, \bibinfo{person}{Meeyoung Cha}, {and} \bibinfo{person}{Pablo Arag\'{o}n}.} \bibinfo{year}{2023}\natexlab{b}.
\newblock \showarticletitle{{A Comparative Study of Reference Reliability in Multiple Language Editions of Wikipedia}}. In \bibinfo{booktitle}{\emph{Proceedings of the 32nd ACM International Conference on Information and Knowledge Management}} \emph{(\bibinfo{series}{CIKM '23})}. \bibinfo{publisher}{ACM}, \bibinfo{address}{New York, NY, USA}, \bibinfo{pages}{3743–3747}.
\newblock
\showISBNx{9798400701245}
\urldef\tempurl%
\url{https://doi.org/10.1145/3583780.3615254}
\showDOI{\tempurl}


\bibitem[\protect\citeauthoryear{Barret}{Barret}{2021}]%
        {iffy}
\bibfield{author}{\bibinfo{person}{Golding Barret}.} \bibinfo{year}{2021}\natexlab{}.
\newblock \bibinfo{title}{Iffy index of unreliable sources}.  (\bibinfo{year}{2021}).
\newblock
\urldef\tempurl%
\url{https://iffy.news/index/}
\showURL{%
\tempurl}
\newblock
\shownote{[accessed 2024 April 3].}


\bibitem[\protect\citeauthoryear{Benjakob, Aviram, and Sobel}{Benjakob et~al\mbox{.}}{2022}]%
        {benjakob2022citation}
\bibfield{author}{\bibinfo{person}{Omer Benjakob}, \bibinfo{person}{Rona Aviram}, {and} \bibinfo{person}{Jonathan~Aryeh Sobel}.} \bibinfo{year}{2022}\natexlab{}.
\newblock \showarticletitle{Citation needed? Wikipedia bibliometrics during the first wave of the COVID-19 pandemic}.
\newblock \bibinfo{journal}{\emph{GigaScience}}  \bibinfo{volume}{11} (\bibinfo{year}{2022}), \bibinfo{pages}{giab095}.
\newblock


\bibitem[\protect\citeauthoryear{Borra, Weltevrede, Ciuccarelli, Kaltenbrunner, Laniado, Magni, Mauri, Rogers, and Venturini}{Borra et~al\mbox{.}}{2015}]%
        {borra2015contropedia}
\bibfield{author}{\bibinfo{person}{Erik Borra}, \bibinfo{person}{Esther Weltevrede}, \bibinfo{person}{Paolo Ciuccarelli}, \bibinfo{person}{Andreas Kaltenbrunner}, \bibinfo{person}{David Laniado}, \bibinfo{person}{Giovanni Magni}, \bibinfo{person}{Michele Mauri}, \bibinfo{person}{Richard Rogers}, {and} \bibinfo{person}{Tommaso Venturini}.} \bibinfo{year}{2015}\natexlab{}.
\newblock \showarticletitle{Societal Controversies in Wikipedia Articles}. In \bibinfo{booktitle}{\emph{Proceedings of the 33rd Annual ACM Conference on Human Factors in Computing Systems}} \emph{(\bibinfo{series}{CHI '15})}. \bibinfo{publisher}{Association for Computing Machinery}, \bibinfo{address}{New York, NY, USA}, \bibinfo{pages}{193–196}.
\newblock
\showISBNx{9781450331456}
\urldef\tempurl%
\url{https://doi.org/10.1145/2702123.2702436}
\showDOI{\tempurl}


\bibitem[\protect\citeauthoryear{Chen and Roth}{Chen and Roth}{2012}]%
        {chen2012citation}
\bibfield{author}{\bibinfo{person}{Chih-Chun Chen} {and} \bibinfo{person}{Camille Roth}.} \bibinfo{year}{2012}\natexlab{}.
\newblock \showarticletitle{$\{$$\{$Citation needed$\}$$\}$ the dynamics of referencing in wikipedia}. In \bibinfo{booktitle}{\emph{Proceedings of the eighth annual international symposium on wikis and open collaboration}}. \bibinfo{publisher}{Association for Computing Machinery}, \bibinfo{address}{New York, NY, USA}, \bibinfo{pages}{1--4}.
\newblock


\bibitem[\protect\citeauthoryear{Chou, Gon{\c{c}}alves, Walton, and Redi}{Chou et~al\mbox{.}}{2020}]%
        {chou2020citation}
\bibfield{author}{\bibinfo{person}{Ai-Jou Chou}, \bibinfo{person}{Guilherme Gon{\c{c}}alves}, \bibinfo{person}{Sam Walton}, {and} \bibinfo{person}{Miriam Redi}.} \bibinfo{year}{2020}\natexlab{}.
\newblock \showarticletitle{Citation detective: a public dataset to improve and quantify wikipedia citation quality at scale}. In \bibinfo{booktitle}{\emph{Proceedings of the Wiki Workshop}}. Wiki Workshop.
\newblock


\bibitem[\protect\citeauthoryear{Chowdhury, Srinivasan, and Getoor}{Chowdhury et~al\mbox{.}}{2020}]%
        {chowdhury2020joint}
\bibfield{author}{\bibinfo{person}{Rajdipa Chowdhury}, \bibinfo{person}{Sriram Srinivasan}, {and} \bibinfo{person}{Lise Getoor}.} \bibinfo{year}{2020}\natexlab{}.
\newblock \showarticletitle{Joint Estimation of User And Publisher Credibility for Fake News Detection}. In \bibinfo{booktitle}{\emph{Proceedings of the 29th ACM International Conference on Information \& Knowledge Management}} \emph{(\bibinfo{series}{CIKM '20})}. \bibinfo{publisher}{Association for Computing Machinery}, \bibinfo{address}{New York, NY, USA}, \bibinfo{pages}{1993–1996}.
\newblock
\showISBNx{9781450368599}
\urldef\tempurl%
\url{https://doi.org/10.1145/3340531.3412066}
\showDOI{\tempurl}


\bibitem[\protect\citeauthoryear{Cohen}{Cohen}{2021}]%
        {wired2021one}
\bibfield{author}{\bibinfo{person}{Noam Cohen}.} \bibinfo{year}{2021}\natexlab{}.
\newblock \bibinfo{title}{{One Woman’s Mission to Rewrite Nazi History on Wikipedia}}.
\newblock
\newblock
\newblock
\shownote{WIRED. [Online; accessed 15-Apr-2024].}


\bibitem[\protect\citeauthoryear{Das, Johnson, Saez-Trumper, and Arag{\'o}n}{Das et~al\mbox{.}}{2024}]%
        {das2024language}
\bibfield{author}{\bibinfo{person}{Paramita Das}, \bibinfo{person}{Isaac Johnson}, \bibinfo{person}{Diego Saez-Trumper}, {and} \bibinfo{person}{Pablo Arag{\'o}n}.} \bibinfo{year}{2024}\natexlab{}.
\newblock \showarticletitle{Language-Agnostic Modeling of Wikipedia Articles for Content Quality Assessment across Languages}.
\newblock \bibinfo{journal}{\emph{Proceedings of the International AAAI Conference on Web and Social Media}} \bibinfo{volume}{18}, \bibinfo{number}{01} (\bibinfo{year}{2024}), \bibinfo{pages}{1--11}.
\newblock


\bibitem[\protect\citeauthoryear{Das, Roy, Chakraborty, and Mukherjee}{Das et~al\mbox{.}}{2025}]%
        {das2024effective}
\bibfield{author}{\bibinfo{person}{Paramita Das}, \bibinfo{person}{Amartya Roy}, \bibinfo{person}{Ritabrata Chakraborty}, {and} \bibinfo{person}{Animesh Mukherjee}.} \bibinfo{year}{2025}\natexlab{}.
\newblock \showarticletitle{On the effective transfer of knowledge from {E}nglish to {H}indi {W}ikipedia}. In \bibinfo{booktitle}{\emph{Proceedings of the 31st International Conference on Computational Linguistics: Industry Track}}. \bibinfo{publisher}{Association for Computational Linguistics}, \bibinfo{address}{Abu Dhabi, UAE}, \bibinfo{pages}{453--465}.
\newblock
\urldef\tempurl%
\url{https://aclanthology.org/2025.coling-industry.39/}
\showURL{%
\tempurl}


\bibitem[\protect\citeauthoryear{Del~Vicario, Bessi, Zollo, Petroni, Scala, Caldarelli, Stanley, and Quattrociocchi}{Del~Vicario et~al\mbox{.}}{2016}]%
        {del2016spreading}
\bibfield{author}{\bibinfo{person}{Michela Del~Vicario}, \bibinfo{person}{Alessandro Bessi}, \bibinfo{person}{Fabiana Zollo}, \bibinfo{person}{Fabio Petroni}, \bibinfo{person}{Antonio Scala}, \bibinfo{person}{Guido Caldarelli}, \bibinfo{person}{H~Eugene Stanley}, {and} \bibinfo{person}{Walter Quattrociocchi}.} \bibinfo{year}{2016}\natexlab{}.
\newblock \showarticletitle{The spreading of misinformation online}.
\newblock \bibinfo{journal}{\emph{Proceedings of the national academy of Sciences}} \bibinfo{volume}{113}, \bibinfo{number}{3} (\bibinfo{year}{2016}), \bibinfo{pages}{554--559}.
\newblock


\bibitem[\protect\citeauthoryear{Devlin, Chang, Lee, and Toutanova}{Devlin et~al\mbox{.}}{2019}]%
        {devlin2018bert}
\bibfield{author}{\bibinfo{person}{Jacob Devlin}, \bibinfo{person}{Ming-Wei Chang}, \bibinfo{person}{Kenton Lee}, {and} \bibinfo{person}{Kristina Toutanova}.} \bibinfo{year}{2019}\natexlab{}.
\newblock \showarticletitle{{BERT}: Pre-training of Deep Bidirectional Transformers for Language Understanding}. In \bibinfo{booktitle}{\emph{Proceedings of the 2019 Conference of the North {A}merican Chapter of the Association for Computational Linguistics: Human Language Technologies, Volume 1 (Long and Short Papers)}}. \bibinfo{publisher}{Association for Computational Linguistics}, \bibinfo{address}{Minneapolis, Minnesota}, \bibinfo{pages}{4171--4186}.
\newblock
\urldef\tempurl%
\url{https://doi.org/10.18653/v1/N19-1423}
\showDOI{\tempurl}


\bibitem[\protect\citeauthoryear{Feith, Arora, Gerlach, Paul, and West}{Feith et~al\mbox{.}}{2024}]%
        {feith2024entity}
\bibfield{author}{\bibinfo{person}{Tom{\'a}s Feith}, \bibinfo{person}{Akhil Arora}, \bibinfo{person}{Martin Gerlach}, \bibinfo{person}{Debjit Paul}, {and} \bibinfo{person}{Robert West}.} \bibinfo{year}{2024}\natexlab{}.
\newblock \showarticletitle{Entity Insertion in Multilingual Linked Corpora: The Case of {W}ikipedia}. In \bibinfo{booktitle}{\emph{Proceedings of the 2024 Conference on Empirical Methods in Natural Language Processing}}. \bibinfo{publisher}{Association for Computational Linguistics}, \bibinfo{address}{Miami, Florida, USA}, \bibinfo{pages}{22796--22819}.
\newblock
\urldef\tempurl%
\url{https://doi.org/10.18653/v1/2024.emnlp-main.1268}
\showDOI{\tempurl}


\bibitem[\protect\citeauthoryear{Fu, Yang, and Fujigaki}{Fu et~al\mbox{.}}{2023}]%
        {fu2023introducing}
\bibfield{author}{\bibinfo{person}{Mengyuan Fu}, \bibinfo{person}{Kunhao Yang}, {and} \bibinfo{person}{Yuko Fujigaki}.} \bibinfo{year}{2023}\natexlab{}.
\newblock \showarticletitle{Introducing an “invisible enemy”: a case study of knowledge construction regarding microplastics in Japanese Wikipedia}.
\newblock \bibinfo{journal}{\emph{New Media \& Society}} \bibinfo{volume}{0}, \bibinfo{number}{0} (\bibinfo{year}{2023}), \bibinfo{pages}{14614448221149747}.
\newblock


\bibitem[\protect\citeauthoryear{Gerlach, Miller, Ho, Harlan, and Difallah}{Gerlach et~al\mbox{.}}{2021}]%
        {gerlach2021multilingual}
\bibfield{author}{\bibinfo{person}{Martin Gerlach}, \bibinfo{person}{Marshall Miller}, \bibinfo{person}{Rita Ho}, \bibinfo{person}{Kosta Harlan}, {and} \bibinfo{person}{Djellel Difallah}.} \bibinfo{year}{2021}\natexlab{}.
\newblock \showarticletitle{Multilingual entity linking system for Wikipedia with a machine-in-the-loop approach}. In \bibinfo{booktitle}{\emph{Proceedings of the 30th ACM International Conference on Information \& Knowledge Management}}. \bibinfo{publisher}{ACM}, \bibinfo{address}{New York, NY, USA}, \bibinfo{pages}{3818--3827}.
\newblock


\bibitem[\protect\citeauthoryear{Guo, Zeng, Tang, and Zhao}{Guo et~al\mbox{.}}{2023b}]%
        {guo2023interpretable}
\bibfield{author}{\bibinfo{person}{Hao Guo}, \bibinfo{person}{Weixin Zeng}, \bibinfo{person}{Jiuyang Tang}, {and} \bibinfo{person}{Xiang Zhao}.} \bibinfo{year}{2023}\natexlab{b}.
\newblock \showarticletitle{Interpretable Fake News Detection with Graph Evidence}. In \bibinfo{booktitle}{\emph{Proceedings of the 32nd ACM International Conference on Information and Knowledge Management}} \emph{(\bibinfo{series}{CIKM '23})}. \bibinfo{publisher}{Association for Computing Machinery}, \bibinfo{address}{New York, NY, USA}, \bibinfo{pages}{659–668}.
\newblock
\showISBNx{9798400701245}
\urldef\tempurl%
\url{https://doi.org/10.1145/3583780.3614936}
\showDOI{\tempurl}


\bibitem[\protect\citeauthoryear{Guo, Han, Lou, Wang, Liu, and Yuan}{Guo et~al\mbox{.}}{2023a}]%
        {guo2023edit}
\bibfield{author}{\bibinfo{person}{Yuhan Guo}, \bibinfo{person}{Qin Han}, \bibinfo{person}{Yuke Lou}, \bibinfo{person}{Yiming Wang}, \bibinfo{person}{Can Liu}, {and} \bibinfo{person}{Xiaoru Yuan}.} \bibinfo{year}{2023}\natexlab{a}.
\newblock \showarticletitle{Edit-History Vis: An Interactive Visual Exploration and Analysis on Wikipedia Edit History}. In \bibinfo{booktitle}{\emph{2023 IEEE 16th Pacific Visualization Symposium (PacificVis)}}. \bibinfo{publisher}{IEEE}, \bibinfo{address}{New York, NY, USA}, \bibinfo{pages}{157--166}.
\newblock


\bibitem[\protect\citeauthoryear{Hale}{Hale}{2014}]%
        {hale2014multilinguals}
\bibfield{author}{\bibinfo{person}{Scott~A. Hale}.} \bibinfo{year}{2014}\natexlab{}.
\newblock \showarticletitle{Multilinguals and Wikipedia editing}. In \bibinfo{booktitle}{\emph{Proceedings of the 2014 ACM Conference on Web Science}} \emph{(\bibinfo{series}{WebSci '14})}. \bibinfo{publisher}{Association for Computing Machinery}, \bibinfo{address}{New York, NY, USA}, \bibinfo{pages}{99–108}.
\newblock
\showISBNx{9781450326223}
\urldef\tempurl%
\url{https://doi.org/10.1145/2615569.2615684}
\showDOI{\tempurl}


\bibitem[\protect\citeauthoryear{Halfaker and Geiger}{Halfaker and Geiger}{2020}]%
        {halfaker2020ores}
\bibfield{author}{\bibinfo{person}{Aaron Halfaker} {and} \bibinfo{person}{R~Stuart Geiger}.} \bibinfo{year}{2020}\natexlab{}.
\newblock \showarticletitle{{ORES: Lowering barriers with participatory machine learning in wikipedia}}.
\newblock \bibinfo{journal}{\emph{Proceedings of the ACM on Human-Computer Interaction}} \bibinfo{volume}{4}, \bibinfo{number}{CSCW2} (\bibinfo{year}{2020}), \bibinfo{pages}{1--37}.
\newblock


\bibitem[\protect\citeauthoryear{Halitaj and Zubiaga}{Halitaj and Zubiaga}{2024}]%
        {halitaj2024providing}
\bibfield{author}{\bibinfo{person}{Aida Halitaj} {and} \bibinfo{person}{Arkaitz Zubiaga}.} \bibinfo{year}{2024}\natexlab{}.
\newblock \showarticletitle{Providing citations to support fact-checking: Contextualizing detection of sentences needing citation on small wikipedias}.
\newblock \bibinfo{journal}{\emph{Natural Language Processing Journal}}  \bibinfo{volume}{8} (\bibinfo{year}{2024}), \bibinfo{pages}{100093}.
\newblock


\bibitem[\protect\citeauthoryear{Halitaj and Zubiaga}{Halitaj and Zubiaga}{2025}]%
        {halitaj2025alpet}
\bibfield{author}{\bibinfo{person}{Aida Halitaj} {and} \bibinfo{person}{Arkaitz Zubiaga}.} \bibinfo{year}{2025}\natexlab{}.
\newblock \showarticletitle{ALPET: Active few-shot learning for citation worthiness detection in low-resource Wikipedia languages}.
\newblock \bibinfo{journal}{\emph{Expert Systems With Applications}}  \bibinfo{volume}{281} (\bibinfo{year}{2025}), \bibinfo{pages}{127503}.
\newblock


\bibitem[\protect\citeauthoryear{Johnson, Gerlach, and S\'{a}ez-Trumper}{Johnson et~al\mbox{.}}{2021}]%
        {johnson2021language}
\bibfield{author}{\bibinfo{person}{Isaac Johnson}, \bibinfo{person}{Martin Gerlach}, {and} \bibinfo{person}{Diego S\'{a}ez-Trumper}.} \bibinfo{year}{2021}\natexlab{}.
\newblock \showarticletitle{Language-agnostic Topic Classification for Wikipedia}. In \bibinfo{booktitle}{\emph{Companion Proceedings of the Web Conference 2021}} \emph{(\bibinfo{series}{WWW '21})}. \bibinfo{publisher}{Association for Computing Machinery}, \bibinfo{address}{New York, NY, USA}, \bibinfo{pages}{594–601}.
\newblock
\showISBNx{9781450383134}
\urldef\tempurl%
\url{https://doi.org/10.1145/3442442.3452347}
\showDOI{\tempurl}


\bibitem[\protect\citeauthoryear{Johnson, Kaffee, and Redi}{Johnson et~al\mbox{.}}{2024}]%
        {johnson2024wikimedia}
\bibfield{author}{\bibinfo{person}{Isaac Johnson}, \bibinfo{person}{Lucie-Aim{\'e}e Kaffee}, {and} \bibinfo{person}{Miriam Redi}.} \bibinfo{year}{2024}\natexlab{}.
\newblock \showarticletitle{Wikimedia data for AI: a review of Wikimedia datasets for NLP tasks and AI-assisted editing}.
\newblock \bibinfo{journal}{\emph{arXiv preprint arXiv:2410.08918}} (\bibinfo{year}{2024}).
\newblock


\bibitem[\protect\citeauthoryear{Johnson and Lescak}{Johnson and Lescak}{2022}]%
        {johnson2022considerations}
\bibfield{author}{\bibinfo{person}{Isaac Johnson} {and} \bibinfo{person}{Emily Lescak}.} \bibinfo{year}{2022}\natexlab{}.
\newblock \showarticletitle{Considerations for multilingual wikipedia research}.
\newblock \bibinfo{journal}{\emph{arXiv preprint arXiv:2204.02483}} (\bibinfo{year}{2022}).
\newblock


\bibitem[\protect\citeauthoryear{Kharazian, Starbird, and Hill}{Kharazian et~al\mbox{.}}{2024}]%
        {kharazian2023governance}
\bibfield{author}{\bibinfo{person}{Zarine Kharazian}, \bibinfo{person}{Kate Starbird}, {and} \bibinfo{person}{Benjamin~Mako Hill}.} \bibinfo{year}{2024}\natexlab{}.
\newblock \showarticletitle{Governance Capture in a Self-Governing Community: A Qualitative Comparison of the Croatian, Serbian, Bosnian, and Serbo-Croatian Wikipedias}.
\newblock \bibinfo{journal}{\emph{Proc. ACM Hum.-Comput. Interact.}} \bibinfo{volume}{8}, \bibinfo{number}{CSCW1}, Article \bibinfo{articleno}{61} (\bibinfo{year}{2024}), \bibinfo{numpages}{26}~pages.
\newblock
\urldef\tempurl%
\url{https://doi.org/10.1145/3637338}
\showDOI{\tempurl}


\bibitem[\protect\citeauthoryear{Korte, Bartsch, Beckmann, {El Baff}, Hamm, and Hecking}{Korte et~al\mbox{.}}{2023}]%
        {korte2023causes}
\bibfield{author}{\bibinfo{person}{Jasper~W. Korte}, \bibinfo{person}{Sabine Bartsch}, \bibinfo{person}{Rasmus Beckmann}, \bibinfo{person}{Roxanne {El Baff}}, \bibinfo{person}{Andreas Hamm}, {and} \bibinfo{person}{Tobias Hecking}.} \bibinfo{year}{2023}\natexlab{}.
\newblock \showarticletitle{From causes to consequences, from chat to crisis. The different climate changes of science and Wikipedia}.
\newblock \bibinfo{journal}{\emph{Environmental Science \& Policy}}  \bibinfo{volume}{148} (\bibinfo{year}{2023}), \bibinfo{pages}{103553}.
\newblock
\showISSN{1462-9011}
\urldef\tempurl%
\url{https://doi.org/10.1016/j.envsci.2023.103553}
\showDOI{\tempurl}


\bibitem[\protect\citeauthoryear{Kumar, West, and Leskovec}{Kumar et~al\mbox{.}}{2016}]%
        {kumar2016disinformation}
\bibfield{author}{\bibinfo{person}{Srijan Kumar}, \bibinfo{person}{Robert West}, {and} \bibinfo{person}{Jure Leskovec}.} \bibinfo{year}{2016}\natexlab{}.
\newblock \showarticletitle{Disinformation on the Web: Impact, Characteristics, and Detection of Wikipedia Hoaxes}. In \bibinfo{booktitle}{\emph{Proceedings of the 25th International Conference on World Wide Web}} \emph{(\bibinfo{series}{WWW '16})}. \bibinfo{publisher}{International World Wide Web Conferences Steering Committee}, \bibinfo{address}{Republic and Canton of Geneva, CHE}, \bibinfo{pages}{591–602}.
\newblock
\showISBNx{9781450341431}
\urldef\tempurl%
\url{https://doi.org/10.1145/2872427.2883085}
\showDOI{\tempurl}


\bibitem[\protect\citeauthoryear{Lewoniewski, W{\k{e}}cel, and Abramowicz}{Lewoniewski et~al\mbox{.}}{2019}]%
        {lewoniewski2019multilingual}
\bibfield{author}{\bibinfo{person}{W{\l}odzimierz Lewoniewski}, \bibinfo{person}{Krzysztof W{\k{e}}cel}, {and} \bibinfo{person}{Witold Abramowicz}.} \bibinfo{year}{2019}\natexlab{}.
\newblock \showarticletitle{Multilingual ranking of Wikipedia articles with quality and popularity assessment in different topics}.
\newblock \bibinfo{journal}{\emph{Computers}} \bibinfo{volume}{8}, \bibinfo{number}{3} (\bibinfo{year}{2019}), \bibinfo{pages}{60}.
\newblock


\bibitem[\protect\citeauthoryear{Lewoniewski, W{\k{e}}cel, and Abramowicz}{Lewoniewski et~al\mbox{.}}{2020}]%
        {lewoniewski2020modeling}
\bibfield{author}{\bibinfo{person}{W{\l}odzimierz Lewoniewski}, \bibinfo{person}{Krzysztof W{\k{e}}cel}, {and} \bibinfo{person}{Witold Abramowicz}.} \bibinfo{year}{2020}\natexlab{}.
\newblock \showarticletitle{Modeling popularity and reliability of sources in multilingual Wikipedia}.
\newblock \bibinfo{journal}{\emph{Information}} \bibinfo{volume}{11}, \bibinfo{number}{5} (\bibinfo{year}{2020}), \bibinfo{pages}{263}.
\newblock


\bibitem[\protect\citeauthoryear{Lundberg and Lee}{Lundberg and Lee}{2017}]%
        {NIPS2017_7062}
\bibfield{author}{\bibinfo{person}{Scott~M Lundberg} {and} \bibinfo{person}{Su-In Lee}.} \bibinfo{year}{2017}\natexlab{}.
\newblock \showarticletitle{A Unified Approach to Interpreting Model Predictions}.
\newblock In \bibinfo{booktitle}{\emph{Advances in Neural Information Processing Systems 30}}, \bibfield{editor}{\bibinfo{person}{I.~Guyon}, \bibinfo{person}{U.~V. Luxburg}, \bibinfo{person}{S.~Bengio}, \bibinfo{person}{H.~Wallach}, \bibinfo{person}{R.~Fergus}, \bibinfo{person}{S.~Vishwanathan}, {and} \bibinfo{person}{R.~Garnett}} (Eds.). \bibinfo{publisher}{Curran Associates, Inc.}, \bibinfo{address}{Red Hook, NY, USA}, \bibinfo{pages}{4765--4774}.
\newblock
\urldef\tempurl%
\url{http://papers.nips.cc/paper/7062-a-unified-approach-to-interpreting-model-predictions.pdf}
\showURL{%
\tempurl}


\bibitem[\protect\citeauthoryear{Piccardi and West}{Piccardi and West}{2021}]%
        {piccardi2021crosslingual}
\bibfield{author}{\bibinfo{person}{Tiziano Piccardi} {and} \bibinfo{person}{Robert West}.} \bibinfo{year}{2021}\natexlab{}.
\newblock \showarticletitle{Crosslingual topic modeling with WikiPDA}. In \bibinfo{booktitle}{\emph{Proceedings of the Web Conference 2021}}. \bibinfo{publisher}{Association for Computing Machinery}, \bibinfo{address}{New York, NY, USA}, \bibinfo{pages}{3032--3041}.
\newblock


\bibitem[\protect\citeauthoryear{Redi, Fetahu, Morgan, and Taraborelli}{Redi et~al\mbox{.}}{2019}]%
        {redi2019citation}
\bibfield{author}{\bibinfo{person}{Miriam Redi}, \bibinfo{person}{Besnik Fetahu}, \bibinfo{person}{Jonathan Morgan}, {and} \bibinfo{person}{Dario Taraborelli}.} \bibinfo{year}{2019}\natexlab{}.
\newblock \showarticletitle{Citation Needed: A Taxonomy and Algorithmic Assessment of Wikipedia's Verifiability}. In \bibinfo{booktitle}{\emph{The World Wide Web Conference}} \emph{(\bibinfo{series}{WWW '19})}. \bibinfo{publisher}{Association for Computing Machinery}, \bibinfo{address}{New York, NY, USA}, \bibinfo{pages}{1567–1578}.
\newblock
\showISBNx{9781450366748}
\urldef\tempurl%
\url{https://doi.org/10.1145/3308558.3313618}
\showDOI{\tempurl}


\bibitem[\protect\citeauthoryear{Rijshouwer, Uitermark, and de~Koster}{Rijshouwer et~al\mbox{.}}{2023}]%
        {rijshouwer2023wikipedia}
\bibfield{author}{\bibinfo{person}{Emiel Rijshouwer}, \bibinfo{person}{Justus Uitermark}, {and} \bibinfo{person}{Willem de Koster}.} \bibinfo{year}{2023}\natexlab{}.
\newblock \showarticletitle{Wikipedia: a self-organizing bureaucracy}.
\newblock \bibinfo{journal}{\emph{Information, Communication \& Society}} \bibinfo{volume}{26}, \bibinfo{number}{7} (\bibinfo{year}{2023}), \bibinfo{pages}{1285--1302}.
\newblock


\bibitem[\protect\citeauthoryear{Rogers, Sendijarevic, et~al\mbox{.}}{Rogers et~al\mbox{.}}{2012}]%
        {rogers2012neutral}
\bibfield{author}{\bibinfo{person}{Richard Rogers}, \bibinfo{person}{Emina Sendijarevic}, {et~al\mbox{.}}} \bibinfo{year}{2012}\natexlab{}.
\newblock \bibinfo{title}{Neutral or national point of view? A comparison of Srebrenica articles across Wikipedia’s language versions}.
\newblock , \bibinfo{numpages}{74}~pages.
\newblock


\bibitem[\protect\citeauthoryear{Shaw and Hill}{Shaw and Hill}{2014}]%
        {shaw2014laboratories}
\bibfield{author}{\bibinfo{person}{Aaron Shaw} {and} \bibinfo{person}{Benjamin~M Hill}.} \bibinfo{year}{2014}\natexlab{}.
\newblock \showarticletitle{Laboratories of oligarchy? How the iron law extends to peer production}.
\newblock \bibinfo{journal}{\emph{Journal of Communication}} \bibinfo{volume}{64}, \bibinfo{number}{2} (\bibinfo{year}{2014}), \bibinfo{pages}{215--238}.
\newblock


\bibitem[\protect\citeauthoryear{Shu, Bernard, and Liu}{Shu et~al\mbox{.}}{2019}]%
        {shu2019studying}
\bibfield{author}{\bibinfo{person}{Kai Shu}, \bibinfo{person}{H.~Russell Bernard}, {and} \bibinfo{person}{Huan Liu}.} \bibinfo{year}{2019}\natexlab{}.
\newblock \showarticletitle{Studying Fake News via Network Analysis: Detection and Mitigation}.
\newblock In \bibinfo{booktitle}{\emph{Emerging Research Challenges and Opportunities in Computational Social Network Analysis and Mining}}, \bibfield{editor}{\bibinfo{person}{Nitin Agarwal}, \bibinfo{person}{Nima Dokoohaki}, {and} \bibinfo{person}{Serpil Tokdemir}} (Eds.). \bibinfo{publisher}{Springer International Publishing}, \bibinfo{address}{Cham}, \bibinfo{pages}{43--65}.
\newblock
\showISBNx{978-3-319-94105-9}
\urldef\tempurl%
\url{https://doi.org/10.1007/978-3-319-94105-9_3}
\showDOI{\tempurl}


\bibitem[\protect\citeauthoryear{Su, Yang, Wu, and Zhang}{Su et~al\mbox{.}}{2023}]%
        {su2023mining}
\bibfield{author}{\bibinfo{person}{Xing Su}, \bibinfo{person}{Jian Yang}, \bibinfo{person}{Jia Wu}, {and} \bibinfo{person}{Yuchen Zhang}.} \bibinfo{year}{2023}\natexlab{}.
\newblock \showarticletitle{Mining user-aware multi-relations for fake news detection in large scale online social networks}. In \bibinfo{booktitle}{\emph{Proceedings of the sixteenth ACM international conference on web search and data mining}}. \bibinfo{pages}{51--59}.
\newblock


\bibitem[\protect\citeauthoryear{Swati, Grobelnik, Mladeni{\'c}, and Grobelnik}{Swati et~al\mbox{.}}{2023}]%
        {swati2023commonsense}
\bibfield{author}{\bibinfo{person}{Swati Swati}, \bibinfo{person}{Adrian~Mladeni{\'c} Grobelnik}, \bibinfo{person}{Dunja Mladeni{\'c}}, {and} \bibinfo{person}{Marko Grobelnik}.} \bibinfo{year}{2023}\natexlab{}.
\newblock \showarticletitle{A commonsense-infused language-agnostic learning framework for enhancing prediction of political bias in multilingual news headlines}.
\newblock \bibinfo{journal}{\emph{Knowledge-Based Systems}}  \bibinfo{volume}{277} (\bibinfo{year}{2023}), \bibinfo{pages}{110838}.
\newblock


\bibitem[\protect\citeauthoryear{Trokhymovych, Aslam, Chou, Baeza-Yates, and Saez-Trumper}{Trokhymovych et~al\mbox{.}}{2023}]%
        {trokhymovych2023fair}
\bibfield{author}{\bibinfo{person}{Mykola Trokhymovych}, \bibinfo{person}{Muniza Aslam}, \bibinfo{person}{Ai-Jou Chou}, \bibinfo{person}{Ricardo Baeza-Yates}, {and} \bibinfo{person}{Diego Saez-Trumper}.} \bibinfo{year}{2023}\natexlab{}.
\newblock \showarticletitle{Fair Multilingual Vandalism Detection System for Wikipedia}. In \bibinfo{booktitle}{\emph{Proceedings of the 29th ACM SIGKDD Conference on Knowledge Discovery and Data Mining}} \emph{(\bibinfo{series}{KDD '23})}. \bibinfo{publisher}{Association for Computing Machinery}, \bibinfo{address}{New York, NY, USA}, \bibinfo{pages}{4981–4990}.
\newblock
\showISBNx{9798400701030}
\urldef\tempurl%
\url{https://doi.org/10.1145/3580305.3599823}
\showDOI{\tempurl}


\bibitem[\protect\citeauthoryear{Trokhymovych and Saez-Trumper}{Trokhymovych and Saez-Trumper}{2021}]%
        {trokhymovych2021wikicheck}
\bibfield{author}{\bibinfo{person}{Mykola Trokhymovych} {and} \bibinfo{person}{Diego Saez-Trumper}.} \bibinfo{year}{2021}\natexlab{}.
\newblock \showarticletitle{WikiCheck: An End-to-end Open Source Automatic Fact-Checking API based on Wikipedia}. In \bibinfo{booktitle}{\emph{Proceedings of the 30th ACM International Conference on Information \& Knowledge Management}} \emph{(\bibinfo{series}{CIKM '21})}. \bibinfo{publisher}{Association for Computing Machinery}, \bibinfo{address}{New York, NY, USA}, \bibinfo{pages}{4155–4164}.
\newblock
\showISBNx{9781450384469}
\urldef\tempurl%
\url{https://doi.org/10.1145/3459637.3481961}
\showDOI{\tempurl}


\bibitem[\protect\citeauthoryear{Vosoughi, Roy, and Aral}{Vosoughi et~al\mbox{.}}{2018}]%
        {vosoughi2018spread}
\bibfield{author}{\bibinfo{person}{Soroush Vosoughi}, \bibinfo{person}{Deb Roy}, {and} \bibinfo{person}{Sinan Aral}.} \bibinfo{year}{2018}\natexlab{}.
\newblock \showarticletitle{The spread of true and false news online}.
\newblock \bibinfo{journal}{\emph{science}} \bibinfo{volume}{359}, \bibinfo{number}{6380} (\bibinfo{year}{2018}), \bibinfo{pages}{1146--1151}.
\newblock


\bibitem[\protect\citeauthoryear{{Wikipedia contributors}}{{Wikipedia contributors}}{2024}]%
        {perennial_sources}
\bibfield{author}{\bibinfo{person}{{Wikipedia contributors}}.} \bibinfo{year}{2024}\natexlab{}.
\newblock \bibinfo{title}{English Wikipedia. 2024. Wikipedia:Reliable sources/Perennial sources}.
\newblock
\newblock
\urldef\tempurl%
\url{https://en.wikipedia.org/wiki/Wikipedia:Reliable_sources/Perennial_sources}
\showURL{%
\tempurl}
\newblock
\shownote{[Online; accessed 12-April-2024].}


\bibitem[\protect\citeauthoryear{Wilkinson, Dumontier, Aalbersberg, Appleton, Axton, Baak, Blomberg, Boiten, da~Silva~Santos, Bourne, et~al\mbox{.}}{Wilkinson et~al\mbox{.}}{2016}]%
        {wilkinson2016fair}
\bibfield{author}{\bibinfo{person}{Mark~D Wilkinson}, \bibinfo{person}{Michel Dumontier}, \bibinfo{person}{IJsbrand~Jan Aalbersberg}, \bibinfo{person}{Gabrielle Appleton}, \bibinfo{person}{Myles Axton}, \bibinfo{person}{Arie Baak}, \bibinfo{person}{Niklas Blomberg}, \bibinfo{person}{Jan-Willem Boiten}, \bibinfo{person}{Luiz~Bonino da Silva~Santos}, \bibinfo{person}{Philip~E Bourne}, {et~al\mbox{.}}} \bibinfo{year}{2016}\natexlab{}.
\newblock \showarticletitle{The FAIR Guiding Principles for scientific data management and stewardship}.
\newblock \bibinfo{journal}{\emph{Scientific data}} \bibinfo{volume}{3}, \bibinfo{number}{1} (\bibinfo{year}{2016}), \bibinfo{pages}{1--9}.
\newblock


\bibitem[\protect\citeauthoryear{Wong, Redi, and Saez-Trumper}{Wong et~al\mbox{.}}{2021}]%
        {wong2021wiki}
\bibfield{author}{\bibinfo{person}{KayYen Wong}, \bibinfo{person}{Miriam Redi}, {and} \bibinfo{person}{Diego Saez-Trumper}.} \bibinfo{year}{2021}\natexlab{}.
\newblock \showarticletitle{Wiki-reliability: A large scale dataset for content reliability on wikipedia}. In \bibinfo{booktitle}{\emph{Proceedings of the 44th International ACM SIGIR Conference on Research and Development in Information Retrieval}}. \bibinfo{publisher}{Association for Computing Machinery}, \bibinfo{address}{New York, NY, USA}, \bibinfo{pages}{2437--2442}.
\newblock


\bibitem[\protect\citeauthoryear{Wu, Li, Deng, Xiong, and Hooi}{Wu et~al\mbox{.}}{2023}]%
        {wu2023prompt}
\bibfield{author}{\bibinfo{person}{Jiaying Wu}, \bibinfo{person}{Shen Li}, \bibinfo{person}{Ailin Deng}, \bibinfo{person}{Miao Xiong}, {and} \bibinfo{person}{Bryan Hooi}.} \bibinfo{year}{2023}\natexlab{}.
\newblock \showarticletitle{Prompt-and-Align: Prompt-Based Social Alignment for Few-Shot Fake News Detection}. In \bibinfo{booktitle}{\emph{Proceedings of the 32nd ACM International Conference on Information and Knowledge Management}} \emph{(\bibinfo{series}{CIKM '23})}. \bibinfo{publisher}{Association for Computing Machinery}, \bibinfo{address}{New York, NY, USA}, \bibinfo{pages}{2726–2736}.
\newblock
\showISBNx{9798400701245}
\urldef\tempurl%
\url{https://doi.org/10.1145/3583780.3615015}
\showDOI{\tempurl}


\bibitem[\protect\citeauthoryear{Xiao, Zhang, Shi, Wang, Naseem, and Hu}{Xiao et~al\mbox{.}}{2024}]%
        {xiao2024msynfd}
\bibfield{author}{\bibinfo{person}{Liang Xiao}, \bibinfo{person}{Qi Zhang}, \bibinfo{person}{Chongyang Shi}, \bibinfo{person}{Shoujin Wang}, \bibinfo{person}{Usman Naseem}, {and} \bibinfo{person}{Liang Hu}.} \bibinfo{year}{2024}\natexlab{}.
\newblock \showarticletitle{MSynFD: Multi-hop Syntax aware Fake News Detection}. In \bibinfo{booktitle}{\emph{Proceedings of the ACM on Web Conference 2024}}. \bibinfo{publisher}{Association for Computing Machinery}, \bibinfo{address}{New York, NY, USA}, \bibinfo{pages}{4128--4137}.
\newblock


\bibitem[\protect\citeauthoryear{Xiao, Guo, Huang, Spolaor, and Cheng}{Xiao et~al\mbox{.}}{2023}]%
        {xiao2023hipo}
\bibfield{author}{\bibinfo{person}{Tianshu Xiao}, \bibinfo{person}{Sichang Guo}, \bibinfo{person}{Jingcheng Huang}, \bibinfo{person}{Riccardo Spolaor}, {and} \bibinfo{person}{Xiuzhen Cheng}.} \bibinfo{year}{2023}\natexlab{}.
\newblock \showarticletitle{HiPo: Detecting Fake News via Historical and Multi-Modal Analyses of Social Media Posts}. In \bibinfo{booktitle}{\emph{Proceedings of the 32nd ACM International Conference on Information and Knowledge Management}} \emph{(\bibinfo{series}{CIKM '23})}. \bibinfo{publisher}{Association for Computing Machinery}, \bibinfo{address}{New York, NY, USA}, \bibinfo{pages}{2805–2815}.
\newblock
\showISBNx{9798400701245}
\urldef\tempurl%
\url{https://doi.org/10.1145/3583780.3614914}
\showDOI{\tempurl}


\bibitem[\protect\citeauthoryear{Xu, Deng, and Zhang}{Xu et~al\mbox{.}}{2022}]%
        {xu2022identifying}
\bibfield{author}{\bibinfo{person}{Xiaofei Xu}, \bibinfo{person}{Ke Deng}, {and} \bibinfo{person}{Xiuzhen Zhang}.} \bibinfo{year}{2022}\natexlab{}.
\newblock \showarticletitle{Identifying cost-effective debunkers for multi-stage fake news mitigation campaigns}. In \bibinfo{booktitle}{\emph{Proceedings of the Fifteenth ACM International Conference on Web Search and Data Mining}}. \bibinfo{publisher}{Association for Computing Machinery}, \bibinfo{address}{New York, NY, USA}, \bibinfo{pages}{1206--1214}.
\newblock


\bibitem[\protect\citeauthoryear{Yang and Colavizza}{Yang and Colavizza}{2024}]%
        {yang2024polarization}
\bibfield{author}{\bibinfo{person}{Puyu Yang} {and} \bibinfo{person}{Giovanni Colavizza}.} \bibinfo{year}{2024}\natexlab{}.
\newblock \showarticletitle{Polarization and reliability of news sources in Wikipedia}.
\newblock \bibinfo{journal}{\emph{Online Information Review}} \bibinfo{volume}{ahead-of-print}, \bibinfo{number}{ahead-of-print} (\bibinfo{year}{2024}), \bibinfo{pages}{1–18}.
\newblock


\bibitem[\protect\citeauthoryear{Ye and Skiena}{Ye and Skiena}{2019}]%
        {ye2019mediarank}
\bibfield{author}{\bibinfo{person}{Junting Ye} {and} \bibinfo{person}{Steven Skiena}.} \bibinfo{year}{2019}\natexlab{}.
\newblock \showarticletitle{MediaRank: Computational Ranking of Online News Sources}. In \bibinfo{booktitle}{\emph{Proceedings of the 25th ACM SIGKDD International Conference on Knowledge Discovery \& Data Mining}} \emph{(\bibinfo{series}{KDD '19})}. \bibinfo{publisher}{Association for Computing Machinery}, \bibinfo{address}{New York, NY, USA}, \bibinfo{pages}{2469–2477}.
\newblock
\showISBNx{9781450362016}
\urldef\tempurl%
\url{https://doi.org/10.1145/3292500.3330709}
\showDOI{\tempurl}


\bibitem[\protect\citeauthoryear{Zandt}{Zandt}{2024}]%
        {mbfc}
\bibfield{author}{\bibinfo{person}{Dave~Van Zandt}.} \bibinfo{year}{2024}\natexlab{}.
\newblock \bibinfo{title}{Media Bias/Fact-Check}.  (\bibinfo{year}{2024}).
\newblock
\urldef\tempurl%
\url{https://mediabiasfactcheck.com/}
\showURL{%
\tempurl}
\newblock
\shownote{[accessed 2024 April 23].}


\bibitem[\protect\citeauthoryear{Zhang, Aljunied, Gao, Chia, and Bing}{Zhang et~al\mbox{.}}{2023a}]%
        {zhang2024m3exam}
\bibfield{author}{\bibinfo{person}{Wenxuan Zhang}, \bibinfo{person}{Mahani Aljunied}, \bibinfo{person}{Chang Gao}, \bibinfo{person}{Yew~Ken Chia}, {and} \bibinfo{person}{Lidong Bing}.} \bibinfo{year}{2023}\natexlab{a}.
\newblock \showarticletitle{M3Exam: A Multilingual, Multimodal, Multilevel Benchmark for Examining Large Language Models}. In \bibinfo{booktitle}{\emph{Advances in Neural Information Processing Systems}}, \bibfield{editor}{\bibinfo{person}{A.~Oh}, \bibinfo{person}{T.~Naumann}, \bibinfo{person}{A.~Globerson}, \bibinfo{person}{K.~Saenko}, \bibinfo{person}{M.~Hardt}, {and} \bibinfo{person}{S.~Levine}} (Eds.), Vol.~\bibinfo{volume}{36}. \bibinfo{publisher}{Curran Associates, Inc.}, \bibinfo{address}{New York}, \bibinfo{pages}{5484--5505}.
\newblock
\urldef\tempurl%
\url{https://proceedings.neurips.cc/paper_files/paper/2023/file/117c5c8622b0d539f74f6d1fb082a2e9-Paper-Datasets_and_Benchmarks.pdf}
\showURL{%
\tempurl}


\bibitem[\protect\citeauthoryear{Zhang, Li, Hauer, Shi, and Kondrak}{Zhang et~al\mbox{.}}{2023b}]%
        {zhang2023don}
\bibfield{author}{\bibinfo{person}{Xiang Zhang}, \bibinfo{person}{Senyu Li}, \bibinfo{person}{Bradley Hauer}, \bibinfo{person}{Ning Shi}, {and} \bibinfo{person}{Grzegorz Kondrak}.} \bibinfo{year}{2023}\natexlab{b}.
\newblock \showarticletitle{Don’t trust ChatGPT when your question is not in English: A study of multilingual abilities and types of LLMs}. In \bibinfo{booktitle}{\emph{Proceedings of the 2023 Conference on Empirical Methods in Natural Language Processing}}. \bibinfo{publisher}{Association for Computational Linguistics}, \bibinfo{address}{Singapore}, \bibinfo{pages}{7915--7927}.
\newblock


\bibitem[\protect\citeauthoryear{Zhao, Da, and Yan}{Zhao et~al\mbox{.}}{2021}]%
        {zhao2021detecting}
\bibfield{author}{\bibinfo{person}{Yuehua Zhao}, \bibinfo{person}{Jingwei Da}, {and} \bibinfo{person}{Jiaqi Yan}.} \bibinfo{year}{2021}\natexlab{}.
\newblock \showarticletitle{Detecting health misinformation in online health communities: Incorporating behavioral features into machine learning based approaches}.
\newblock \bibinfo{journal}{\emph{Information Processing \& Management}} \bibinfo{volume}{58}, \bibinfo{number}{1} (\bibinfo{year}{2021}), \bibinfo{pages}{102390}.
\newblock


\bibitem[\protect\citeauthoryear{Zheng, Chen, Yan, and Ni}{Zheng et~al\mbox{.}}{2023}]%
        {zheng2023gender}
\bibfield{author}{\bibinfo{person}{Xiang Zheng}, \bibinfo{person}{Jiajing Chen}, \bibinfo{person}{Erjia Yan}, {and} \bibinfo{person}{Chaoqun Ni}.} \bibinfo{year}{2023}\natexlab{}.
\newblock \showarticletitle{Gender and country biases in Wikipedia citations to scholarly publications}.
\newblock \bibinfo{journal}{\emph{Journal of the Association for Information Science and Technology}} \bibinfo{volume}{74}, \bibinfo{number}{2} (\bibinfo{year}{2023}), \bibinfo{pages}{219--233}.
\newblock


\end{thebibliography}

\appendix
\setcounter{table}{0}
\renewcommand{\thetable}{A\arabic{table}}

\section{Appendix A: Feature list}
\label{appendix_feature_list}
We provide here a more detailed list of the 52 features used by our models.

\subsection{Popularity features}
\vspace{-3pt}

\begin{enumerate}[leftmargin=*]
\item $N_{\mathrm{articles}}$: Number of articles a domain has appeared in (sensitive to the size of the dataset).

\item $\overline{N_{articles}}$: Number of articles a domain has appeared in, normalized by the number of articles in the dataset.

\item $CurrN_{articles}$: Number of articles a domain is used in at collection time.

\item $\overline{CurrN_{articles}}$: Number of articles a domain is used in at collection time, normalized by the total number of articles in the dataset.

\end{enumerate}

\subsection{Permanence features}

\begin{enumerate}[leftmargin=*]
\vspace{-3pt}
  \setcounter{enumi}{4}
\item $\Sigma Perm_d$: Permanence, how long a domain has been used in an article (not necessarily consecutively), summed across all articles and measured in days.

\item $\Sigma Perm_r$: Permanence, how long a domain has been used in an article (not necessarily consecutively), summed across all articles and measured by his number of revisions.

\item $\Sigma CurrPerm_d$: How long a domain has been used in an article (not necessarily consecutively), summed across all articles and measured in days, but considering only the articles where the domain is currently used.

\item $\Sigma CurrPerm_r$: How long a domain has been used in an article (not necessarily consecutively), summed across all articles and measured in number of revisions, but considering only the articles where the domain is currently used.

\item $\overline{\Sigma Perm_d}$: Sum of all permanences for a domain in all articles, normalized by the sum of the ages of all articles, measured by the number of days.

\item $\overline{\Sigma Perm_r}$: Sum of all permanences for a domain in all articles, normalized by the sum of the ages of all articles, measured by the number of revisions.

\item $\overline{\Sigma CurrPerm_d}$: Sum of all permanences for a domain in articles where the domain was used at collection time, normalized by the sum of the ages of all articles, measured by the number of days.

\item $\overline{\Sigma CurrPerm_r}$: Sum of all permanences for a domain in articles where the domain was used at collection time, normalized by the sum of the ages of all articles, measured by the number of revisions.

\item $\langle Perm_d \rangle$: The average permanence of a domain (over all articles where it was used), measured by the number of days.

\item $\langle Perm_r \rangle$: The average permanence of a domain (over all articles where it was used), measured by the number of revisions.

\item $\langle SelfPerm_d\rangle$, : Self-permanence: permanence divided by the number of days since the first time the domains were added to the article, then averaged article-wise.

\item $\langle SelfPerm_r\rangle$: Self-permanence: permanence divided by the number of revisions since the first time the domains were added to the article, then averaged article-wise.

\item $ \Sigma age_d $: The sum of ages of the domain over all the articles in a dataset, where the age of a domain in an article is the number of days since a URL from that domain has been first added to the article.

\item $ \Sigma age_r $: The sum of ages of the domain over all the articles in a dataset, where the age of a domain in an article is the number of revisions since a URL from that domain has been first added to the article.

\item $\langle age_d \rangle$: The average age of a domain over all the articles it appears at least once, measured by the number of days.

\item $\langle age_r \rangle$: The average age of a domain over all the articles it appears at least once, measured by the number of revisions.
\end{enumerate}

\subsection{User-based features}
\begin{enumerate}[leftmargin=*]
\vspace{-3pt}
  \setcounter{enumi}{20}
\item $U_{add}$ Number of unique users that have added a domain.
\item $U_{start}$ Number of unique users that have added a domain for the first time on a page.

\item $U_{rem}$ Number of unique users that have removed a domain.
\item $U_{end}$ Number of unique users that have removed a domain for the last time on a page.

\item $R_{add}$ Number of unique registered users that have added a domain.
\item $R_{start}$ Number of unique registered users that have added a domain for the first time on a page.

\item $R_{rem}$ Number of unique registered users that have removed a domain.
\item $R_{end}$ Number of unique registered users that have removed a domain for the last time on a page.

\item $\overline{U_{add}}$ Number of unique users that have added a domain, divided by the total number of unique users in the dataset.
\item $\overline{U_{start}}$ Number of unique users that have added a domain for the first time on a page, divided by the total number of unique users in the dataset.

\item $\overline{U_{rem}}$: Number of unique users that have removed a domain, divided by the total number of unique users in the dataset.
\item $\overline{U_{end}}$: Number of unique users that have removed a domain for the last time on a page, divided by the total number of unique users in the dataset.

\item $\overline{R_{add}}$: Number of unique registered users that have added a domain, divided by the total number of unique users in the dataset.
\item $\overline{R_{start}}$: Number of unique registered users that have added a domain for the first time on a page, divided by the total number of unique users in the dataset.

\item $\overline{R_{rem}}$: Number of unique registered users that have removed a domain, divided by the total number of unique users in the dataset.
\item $\overline{R_{end}}$: Number of unique registered users that have removed a domain for the last time on a page, divided by the total number of unique users in the dataset.

\item $\langle {U_{add}} \rangle$: Average per page number of unique users that have added a domain.
\item $\langle {U_{start}} \rangle$: Average per page number of unique users that have added a domain for the first time on a page.

\item $\langle {U_{rem}} \rangle$: Average per page number of unique users that have removed a domain.
\item $\langle {U_{end}} \rangle$: Average per page number of unique users that have removed a domain for the last time on a page.

\item $\langle {R_{add}} \rangle$: Average per page number of unique registered users that have added a domain.
\item $\langle {R_{start}} \rangle$: Average per page number of unique registered users that have added a domain for the first time on a page.

\item $\langle {R_{rem}} \rangle$: Average per page number of unique registered users that have removed a domain
\item $\langle {R_{end}} \rangle$: Average per page number of unique registered users that have removed a domain for the last time on a page.

\item $Ratio(R_{add}/U_{add})$: The proportion of registered vs. all users that ever added a domain.
\item $Ratio(R_{start}/U_{start})$: The proportion of registered vs. all users that ever added a domain for the first time on a page.

\item $Ratio(R_{rem}/U_{rem})$: The proportion of registered vs. all users that ever removed a domain.
\item $Ratio(R_{end}/U_{end})$: The proportion of registered vs. all users that ever removed a domain for the last time on a page.

\item $Proba(R_{add})$: Probability that, when a domain is added, this is done by the registered users.
\item $Proba(R_{start})$: Probability that, when a domain is added for the first time on a page, this is done by the registered users.
\item $Proba(R_{rem})$: Probability that, when a domain is removed, this is done by the registered users.
\item $Proba(R_{end})$: Probability that, when a domain is removed for the last time on a page, this is done by the registered users.
\end{enumerate}

\newpage

\section{Appendix B: Medians of resource-wise statistics}
\label{app:medians}

\begin{table*}[h]
\caption{Statistics of the topic-language datasets used in the study grouped by high, mid, and low resource languages. All numbers are medians (apart from the number of languages). Perennial Domains (P. Domains) refers to the number of domains for the perennial source list in English found in the different datasets. The corresponding mean values can be found in Table~\ref{tab:dataset_stats}.}
\label{tab:dataset_stats_median}
\centering
\footnotesize
\begin{tabular}{@{}clrrrrrrrrr@{}}
\toprule
\textbf{Topic} & \textbf{Langs} & \textbf{Articles} & \textbf{Revs} & \textbf{Revs/Ar} & \textbf{URLs} & \textbf{URLs/Ar} & \textbf{Domains} & \textbf{P. Domains} \\
\midrule
\multirow{5}{*}{\rotatebox[origin=c]{90}{High}}
Climate change &       7 &    1162 &  251085 &             62 &  34352 &              18 &  11992 &        191 \\
COVID-19       &       7 &     595 &  136470 &             49 &  29826 &              21 &   6031 &        185 \\
Biology &       7 &    8228 &  388729 &             17 &  32315 &               3 &   8377 &        127 \\
History &       7 &    4839 &  417275 &             40 &  30957 &               5 &   9324 &        146 \\
Media   &       7 &    3542 &  398138 &             46 &  43849 &               8 &  12441 &        240 \\
\midrule
\multirow{5}{*}{\rotatebox[origin=c]{90}{Mid}}
Climate change &      36 &     325 &  23000 &             27 &   4399 &               9 &  1991 &         85 \\
COVID-19       &      37 &     182 &  10026 &             19 &   4891 &               9 &  1326 &         99 \\
Biology &      37 &    2053 &  57234 &             11 &   5005 &               2 &  1643 &         59 \\
History &      37 &    1687 &  70988 &             26 &   5500 &               3 &  1901 &         62 \\
Media   &      37 &     799 &  30098 &             20 &   5347 &               4 &  1991 &        123 \\
\midrule
\multirow{5}{*}{\rotatebox[origin=c]{90}{Low}}
Climate change &      56 &     100 &  2403 &             12 &    633 &               6 &   373 &         25 \\
COVID-19       &      75 &      24 &   518 &              9 &    466 &               9 &   214 &         31 \\
Biology &      41 &     440 &  4949 &              7 &    541 &               3 &   301 &         14 \\
History &      55 &     421 &  8020 &             13 &    449 &               3 &   256 &         16 \\
Media   &      70 &      73 &  1063 &              6 &    263 &               3 &   148 &         22 \\
\bottomrule
\end{tabular}
\end{table*}

\section{APPENDIX C: Language-specific dataset sizes}
\label{app:datasetsizes}

\begin{longtable}{@{}
lrrrrr@{}}
\caption{Number of articles per language editions in the different datasets.} \\
\toprule
Language &  Climate change &  COVID-19 &  Biology &  History &  Media \\
\midrule
\endfirsthead
\toprule
  Language &  Climate change &  COVID-19 &  Biology &  History &  Media \\
\midrule
\endhead
de &             479 &       274 &             463 &             549 &           472 \\
en &             975 &       856 &             931 &            1092 &          1460 \\
es &             339 &       258 &             397 &             461 &           509 \\
fr &             412 &       262 &             388 &             546 &           476 \\
ja &             217 &       145 &             230 &             246 &           382 \\
pt &             223 &       125 &             248 &             258 &           296 \\
ru &             257 &       162 &             317 &             435 &           377 \\
\midrule
ar     &             195 &       148 &             196 &             258 &           189 \\
az     &              62 &        43 &              88 &             126 &            73 \\
bg     &             109 &        44 &             141 &             175 &           116 \\
bn     &              82 &        51 &              92 &              95 &            80 \\
ca     &             176 &        94 &             217 &             242 &           148 \\
cs     &             168 &        79 &             188 &             203 &           143 \\
da     &             141 &        43 &             143 &             161 &           119 \\
el     &             104 &        55 &             112 &             211 &            95 \\
et     &             101 &        58 &             105 &             122 &            73 \\
eu     &             104 &        36 &              94 &             144 &            68 \\
fa     &             156 &        89 &             171 &             229 &           168 \\
fi     &             184 &        68 &             184 &             177 &           178 \\
he     &             184 &       146 &             230 &             301 &           225 \\
hi     &              98 &        42 &              78 &              76 &            58 \\
hr     &              99 &        33 &             104 &             148 &            83 \\
hu     &             151 &        61 &             205 &             227 &           186 \\
hy     &              56 &        39 &              70 &             112 &            83 \\
id     &             136 &        83 &             129 &             154 &           122 \\
it     &             259 &       161 &             289 &             426 &           427 \\
ko     &             139 &       106 &             153 &             204 &           204 \\
ms     &              73 &        63 &              82 &              91 &            67 \\
nb     &             174 &        72 &             170 &             183 &           145 \\
nl     &             238 &       114 &             270 &             295 &           233 \\
pl     &             185 &        89 &             233 &             279 &           236 \\
ro     &             106 &        45 &             122 &             171 &           109 \\
simple &             136 &        53 &             140 &             137 &           118 \\
sk     &             105 &        26 &             118 &             137 &            91 \\
sl     &             106 &        39 &              99 &             156 &            65 \\
sr     &             115 &        59 &             129 &             230 &           126 \\
sv     &             187 &        79 &             204 &             221 &           201 \\
th     &             107 &        61 &             123 &             138 &           125 \\
tr     &             157 &        89 &             160 &             247 &           182 \\
uk     &             162 &       212 &             173 &             194 &           150 \\
uz     &              36 &        39 &              33 &              84 &            42 \\
vi     &             137 &       217 &             145 &             161 &           145 \\
zh     &             205 &       246 &             212 &             250 &           326 \\
kk     &               0 &        17 &              55 &              49 &            34 \\
\midrule
af         &              72 &        28 &              71 &             102 &            46 \\
als        &              33 &         0 &               0 &              58 &             0 \\
as         &              21 &        13 &              23 &              18 &            13 \\
ast        &              84 &        20 &              88 &             130 &            84 \\
ba         &              28 &         6 &               0 &              36 &            14 \\
bcl        &              21 &         9 &              16 &              27 &            12 \\
be         &              56 &        27 &              61 &              90 &            48 \\
be-x-old   &              45 &        14 &               0 &              77 &            42 \\
bs         &              74 &        21 &              71 &             118 &            52 \\
ckb        &              31 &        16 &               0 &              44 &            38 \\
cv         &              25 &         0 &               0 &               0 &            25 \\
cy         &              63 &        23 &             556 &             161 &            55 \\
dag        &              29 &         0 &               0 &               0 &             0 \\
eo         &             104 &        26 &             121 &             137 &            79 \\
ga         &              43 &        24 &              44 &              76 &            47 \\
gl         &             108 &        34 &             127 &             152 &            81 \\
gu         &              29 &         6 &               0 &               0 &             9 \\
ha         &              32 &        23 &               0 &              20 &             8 \\
ia         &              31 &        11 &               0 &               0 &             0 \\
ig         &              17 &         0 &               7 &              10 &             9 \\
jv         &              46 &        15 &              63 &              74 &            32 \\
ka         &              66 &        24 &              76 &             113 &            77 \\
kn         &              51 &        28 &              40 &              33 &            35 \\
ku         &              30 &        13 &               0 &              58 &            31 \\
la         &              69 &        19 &              89 &             130 &            54 \\
lb         &              33 &         0 &               0 &               0 &            53 \\
lt         &             102 &        26 &             120 &             129 &            79 \\
lv         &              80 &        27 &              95 &              93 &            63 \\
mk         &              72 &        18 &              73 &             124 &            59 \\
ml         &              70 &        39 &              83 &              81 &            49 \\
mn         &              43 &         0 &              42 &               0 &            18 \\
mni        &              13 &         0 &               0 &               0 &             0 \\
mnw        &               4 &         5 &               0 &               0 &             0 \\
mr         &              63 &        20 &               0 &               0 &            33 \\
my         &              37 &        18 &               0 &              29 &            16 \\
ne         &              35 &        16 &              42 &               0 &            23 \\
nn         &              87 &        14 &              83 &             101 &            59 \\
oc         &              55 &         0 &               0 &              98 &            38 \\
or         &              16 &        18 &               0 &               0 &            14 \\
pa         &              34 &        24 &              29 &              36 &            25 \\
pnb        &              28 &         9 &              38 &              38 &            16 \\
ps         &              21 &         9 &               0 &              23 &             0 \\
rm         &               6 &         0 &               0 &               0 &             0 \\
rw         &              14 &         0 &              19 &               0 &             0 \\
sh         &              67 &        20 &              62 &             100 &            52 \\
si         &              51 &        16 &              32 &              24 &            16 \\
sq         &              56 &        36 &              50 &             104 &            60 \\
su         &              30 &        19 &              44 &               0 &            13 \\
sw         &              44 &        14 &              47 &              74 &            31 \\
ta         &              90 &        33 &              88 &              68 &            57 \\
te         &              49 &        33 &              61 &              40 &            47 \\
tg         &              30 &         9 &               0 &               0 &            23 \\
tl         &              47 &        40 &              65 &              76 &            41 \\
ur         &              57 &        37 &              58 &              99 &            42 \\
war        &              34 &         0 &               0 &              48 &            17 \\
zh-yue     &              58 &        54 &              62 &              59 &            49 \\
am         &               0 &         6 &               0 &               0 &             0 \\
ary        &               0 &         4 &               0 &               0 &             0 \\
arz        &               0 &        23 &              32 &              79 &            38 \\
awa        &               0 &         2 &               0 &               0 &             0 \\
bjn        &               0 &        11 &               0 &               0 &             0 \\
bo         &               0 &         4 &               0 &               0 &             0 \\
br         &               0 &        13 &               0 &              90 &            41 \\
bug        &               0 &         2 &               0 &               0 &             0 \\
bxr        &               0 &         7 &               0 &               0 &             0 \\
ceb        &               0 &         7 &               0 &               0 &             0 \\
crh        &               0 &         3 &               0 &               0 &             0 \\
dty        &               0 &         2 &               0 &               0 &             0 \\
gn         &               0 &         6 &               0 &               0 &             0 \\
gpe        &               0 &         1 &               0 &               0 &             0 \\
ht         &               0 &        26 &               0 &               0 &             0 \\
hyw        &               0 &         3 &               0 &              33 &            18 \\
km         &               0 &         8 &               0 &               9 &             0 \\
mg         &               0 &         7 &               0 &               0 &             0 \\
min        &               0 &        17 &               0 &               0 &             0 \\
pap        &               0 &        12 &               0 &               0 &             0 \\
sco        &               0 &        13 &              31 &              31 &            24 \\
shn        &               0 &         5 &               0 &               0 &             0 \\
so         &               0 &         9 &               0 &               0 &             0 \\
tk         &               0 &         6 &               0 &               0 &             0 \\
ts         &               0 &         5 &               0 &               0 &             0 \\
vo         &               0 &         3 &               0 &               0 &            12 \\
wuu        &               0 &        15 &               0 &               0 &             0 \\
xmf        &               0 &        10 &               0 &               0 &            17 \\
yo         &               0 &        24 &               0 &               0 &             0 \\
zh-min-nan &               0 &        14 &               0 &              39 &            25 \\
is         &               0 &         0 &              75 &              91 &            50 \\
ky         &               0 &         0 &              21 &              20 &            12 \\
pam        &               0 &         0 &              27 &               0 &             0 \\
tt         &               0 &         0 &              26 &              57 &            14 \\
azb        &               0 &         0 &               0 &              32 &            12 \\
fo         &               0 &         0 &               0 &              26 &            16 \\
fy         &               0 &         0 &               0 &              81 &            32 \\
io         &               0 &         0 &               0 &              79 &             0 \\
tum        &               0 &         0 &               0 &               2 &             0 \\
an         &               0 &         0 &               0 &               0 &            35 \\
frp        &               0 &         0 &               0 &               0 &             6 \\
glk        &               0 &         0 &               0 &               0 &             3 \\
lo         &               0 &         0 &               0 &               0 &             5 \\
mai        &               0 &         0 &               0 &               0 &             8 \\
nds        &               0 &         0 &               0 &               0 &            28 \\
qu         &               0 &         0 &               0 &               0 &            15 \\
sat        &               0 &         0 &               0 &               0 &            12 \\
vec        &               0 &         0 &               0 &               0 &            19 \\

\bottomrule
\end{longtable}

\newpage


\begin{longtable}{lrrrrr}
\caption{Number of revisions per language editions in the different datasets.} \\
\toprule
Language &  Climate change &  COVID-19 &  Biology &  History &  Media \\
\midrule
\endfirsthead
\toprule
Language &  Climate change &  COVID-19 &  Biology &  History &  Media \\
\midrule
\endhead

de &          344149 &    195448 &          388729 &          549521 &        398138 \\
en &         1516501 &   1036268 &         2392272 &         2213972 &       3850137 \\
es &          251085 &    142608 &          412920 &          417275 &        447902 \\
fr &          261336 &    136470 &          401974 &          570207 &        405992 \\
ja &           59235 &     37293 &           97684 &          117733 &        222585 \\
pt &           82137 &     32740 &          178778 &          153532 &        154532 \\
ru &           90492 &     56512 &          230846 &          348758 &        257507 \\
\midrule
ar     &           64147 &     52221 &          123245 &          164152 &         73088 \\
az     &            5396 &      3275 &           37635 &           34400 &          6289 \\
bg     &           14688 &      4606 &           50747 &           77809 &         24812 \\
bn     &            9138 &      6567 &           12759 &           20227 &          7907 \\
ca     &           43861 &     13878 &          219014 &          168055 &         45265 \\
cs     &           37535 &     15008 &           70480 &           94004 &         36957 \\
da     &           21730 &      4267 &           33759 &           54452 &         21325 \\
el     &           15873 &      7175 &           16066 &           93809 &         12233 \\
et     &           12000 &      6702 &           21939 &           28705 &          6906 \\
eu     &           12238 &      2086 &           82623 &           45030 &          6865 \\
fa     &           34394 &     14882 &           65428 &           89813 &         59550 \\
fi     &           41846 &     10081 &           76296 &           70988 &         58117 \\
he     &           45760 &     36839 &           84560 &          153974 &         74220 \\
hi     &           10574 &      3708 &            7683 &           12406 &          4471 \\
hr     &           11084 &      2842 &           21316 &           40023 &          9546 \\
hu     &           27013 &     14401 &           93431 &          100103 &         50952 \\
hy     &            5149 &      4177 &           10735 &           29745 &         12491 \\
id     &           23373 &     14547 &          111305 &           55091 &         32762 \\
it     &          102487 &     46926 &          204053 &          375945 &        347881 \\
ko     &           23966 &     28724 &           57234 &           84491 &         88325 \\
ms     &            6906 &      7698 &           11679 &           20204 &          7208 \\
nb     &           36680 &     10026 &           81352 &           78627 &         40317 \\
nl     &           78355 &     25913 &          303604 &          171942 &         93987 \\
pl     &           45284 &     22540 &          160392 &          173581 &        116880 \\
ro     &           14046 &      7347 &           25254 &           60565 &         19300 \\
simple &           30367 &      8317 &           31239 &           33583 &         23031 \\
sk     &           11945 &      2913 &           18615 &           40322 &          9525 \\
sl     &           12320 &      2265 &           12780 &           39772 &          5602 \\
sr     &           14728 &      6756 &           46083 &          102224 &         20365 \\
sv     &           49949 &     11079 &          264477 &          114709 &         71292 \\
th     &           12084 &     11217 &           22152 &           36162 &         30098 \\
tr     &           34878 &     19488 &           61404 &          121987 &         56558 \\
uk     &           33040 &     55024 &          105302 &          102155 &         45483 \\
uz     &            2067 &      3116 &            2961 &           17800 &          2682 \\
vi     &           22628 &     66920 &          407504 &           45861 &         30883 \\
zh     &           53587 &    131339 &          131500 &          122649 &        182721 \\
kk     &               0 &       517 &            6564 &            5187 &          1523 \\
\midrule
af         &            6384 &      2827 &           12553 &           20980 &          3448 \\
als        &            1320 &         0 &               0 &            4863 &             0 \\
as         &             814 &       466 &            1025 &             590 &           270 \\
ast        &            7812 &       997 &           40421 &           36452 &          8628 \\
ba         &            1316 &        50 &               0 &            2004 &           449 \\
bcl        &            1036 &       182 &             439 &            1082 &           158 \\
be         &            3892 &      1473 &           10090 &           23364 &          2899 \\
be-x-old   &            2385 &       518 &               0 &           12592 &          1722 \\
bs         &            5029 &      1324 &            7127 &           30720 &          2982 \\
ckb        &            1397 &       564 &               0 &            3609 &          2026 \\
cv         &             860 &         0 &               0 &               0 &           746 \\
cy         &            4720 &      1280 &          521785 &           50767 &          9683 \\
dag        &             790 &         0 &               0 &               0 &             0 \\
eo         &           11720 &      1920 &           29807 &           42486 &          7768 \\
ga         &            2570 &      1830 &            4293 &            8020 &          2404 \\
gl         &           13814 &      2608 &           32768 &           44564 &         11127 \\
gu         &            1191 &       311 &               0 &               0 &           157 \\
ha         &            1676 &       638 &               0 &             628 &           131 \\
ia         &            1437 &       265 &               0 &               0 &             0 \\
ig         &             884 &         0 &             116 &             284 &           166 \\
jv         &            2325 &       492 &            4949 &           12716 &          1271 \\
ka         &            5100 &      1312 &           11997 &           33511 &          8342 \\
kn         &            2643 &      1366 &            2085 &            1482 &          1593 \\
ku         &            1659 &       656 &               0 &            4968 &          1020 \\
la         &            5306 &       749 &           11783 &           40788 &          3739 \\
lb         &            1925 &         0 &               0 &               0 &          2999 \\
lt         &           11336 &      1616 &           32199 &           34848 &          7025 \\
lv         &            7180 &      4234 &           13517 &           12402 &          5800 \\
mk         &            6096 &       811 &            8133 &           35526 &          4200 \\
ml         &            6461 &      2373 &           14983 &            7912 &          3545 \\
mn         &            2421 &         0 &            2011 &               0 &           474 \\
mni        &             385 &         0 &               0 &               0 &             0 \\
mnw        &             160 &       217 &               0 &               0 &             0 \\
mr         &            4801 &      1140 &               0 &               0 &          1494 \\
my         &            1469 &      1358 &               0 &            1317 &           430 \\
ne         &            1383 &       487 &            3090 &               0 &           684 \\
nn         &            8312 &      1801 &           10620 &           20234 &          4729 \\
oc         &            4226 &         0 &               0 &           19691 &          1618 \\
or         &             313 &       916 &               0 &               0 &           286 \\
pa         &            1609 &       825 &            1498 &            2143 &           832 \\
pnb        &            1076 &       100 &            4884 &            4351 &           290 \\
ps         &             693 &       144 &               0 &             850 &             0 \\
rm         &             148 &         0 &               0 &               0 &             0 \\
rw         &             390 &         0 &             400 &               0 &             0 \\
sh         &            4436 &       619 &           13554 &           28448 &          5352 \\
si         &            2728 &       421 &            1249 &             869 &           469 \\
sq         &            3919 &      2758 &            3169 &           24390 &          4022 \\
su         &            1328 &       523 &            1885 &               0 &           389 \\
sw         &            3057 &       416 &            2273 &           13290 &          1027 \\
ta         &           10484 &      2321 &           12039 &            6974 &          4530 \\
te         &            3599 &      3142 &            4649 &            2462 &          3206 \\
tg         &            1414 &       331 &               0 &               0 &           637 \\
tl         &            3262 &      2685 &            6835 &           11473 &          2215 \\
ur         &            4132 &      2055 &            3941 &           18574 &          2528 \\
war        &            1436 &         0 &               0 &            7500 &           419 \\
zh-yue     &            3844 &      6256 &            6081 &            7762 &          3438 \\
am         &               0 &       115 &               0 &               0 &             0 \\
ary        &               0 &       539 &               0 &               0 &             0 \\
arz        &               0 &      1221 &           12611 &           18006 &         10561 \\
awa        &               0 &        48 &               0 &               0 &             0 \\
bjn        &               0 &       153 &               0 &               0 &             0 \\
bo         &               0 &        57 &               0 &               0 &             0 \\
br         &               0 &       530 &               0 &           22230 &          2022 \\
bug        &               0 &         7 &               0 &               0 &             0 \\
bxr        &               0 &       238 &               0 &               0 &             0 \\
ceb        &               0 &       125 &               0 &               0 &             0 \\
crh        &               0 &        50 &               0 &               0 &             0 \\
dty        &               0 &        16 &               0 &               0 &             0 \\
gn         &               0 &        89 &               0 &               0 &             0 \\
gpe        &               0 &        31 &               0 &               0 &             0 \\
ht         &               0 &      1304 &               0 &               0 &             0 \\
hyw        &               0 &       118 &               0 &            1401 &           401 \\
km         &               0 &       464 &               0 &             134 &             0 \\
mg         &               0 &       138 &               0 &               0 &             0 \\
min        &               0 &       518 &               0 &               0 &             0 \\
pap        &               0 &       387 &               0 &               0 &             0 \\
sco        &               0 &       223 &            1636 &            1609 &           689 \\
shn        &               0 &       143 &               0 &               0 &             0 \\
so         &               0 &       164 &               0 &               0 &             0 \\
tk         &               0 &        95 &               0 &               0 &             0 \\
ts         &               0 &        98 &               0 &               0 &             0 \\
vo         &               0 &        33 &               0 &               0 &           412 \\
wuu        &               0 &       341 &               0 &               0 &             0 \\
xmf        &               0 &       370 &               0 &               0 &           342 \\
yo         &               0 &      1353 &               0 &               0 &             0 \\
zh-min-nan &               0 &       393 &               0 &            3607 &           889 \\
is         &               0 &         0 &            7778 &           12301 &          2541 \\
ky         &               0 &         0 &            1317 &             705 &           297 \\
pam        &               0 &         0 &             798 &               0 &             0 \\
tt         &               0 &         0 &            2568 &           11215 &           459 \\
azb        &               0 &         0 &               0 &            3732 &          1496 \\
fo         &               0 &         0 &               0 &            1252 &           307 \\
fy         &               0 &         0 &               0 &           12889 &          1099 \\
io         &               0 &         0 &               0 &           15974 &             0 \\
tum        &               0 &         0 &               0 &              16 &             0 \\
an         &               0 &         0 &               0 &               0 &          1325 \\
frp        &               0 &         0 &               0 &               0 &            36 \\
glk        &               0 &         0 &               0 &               0 &            61 \\
lo         &               0 &         0 &               0 &               0 &            47 \\
mai        &               0 &         0 &               0 &               0 &           104 \\
nds        &               0 &         0 &               0 &               0 &           801 \\
qu         &               0 &         0 &               0 &               0 &           435 \\
sat        &               0 &         0 &               0 &               0 &           133 \\
vec        &               0 &         0 &               0 &               0 &           569 \\
\bottomrule
\end{longtable}

\end{document}